\documentclass[12pt]{article}
\usepackage{amssymb}
\usepackage{geometry}
\usepackage{amsmath}
\usepackage{html}
\usepackage{fancyvrb,xspace}
\usepackage{float}
\usepackage{scalefnt}
\usepackage[none]{hyphenat}
\usepackage{multicol}
\usepackage{subfigure}
\usepackage{a4}
\usepackage{longtable}
\usepackage{lscape}
\usepackage{verbatim}
\usepackage{graphicx}
\usepackage{graphics}
\usepackage{harvard}
\usepackage{wrapfig}
\usepackage[onehalfspacing]{setspace}
\usepackage{array}
\usepackage{dcolumn}
\usepackage{fancyhdr}
\usepackage{lastpage}
\usepackage[draft]{fixme}
\usepackage{dcolumn}
\usepackage{setspace}
\usepackage[draft]{fixme}
\usepackage{rotating}
\usepackage{changepage}
\usepackage{multirow}
\usepackage{placeins}
\usepackage{pgfplots}
\usepackage{adjustbox}
\usepackage{tikz}
\usepackage{tkz-graph}
\usepackage[textsize=scriptsize, ]{todonotes}
\usepackage{bm}

\pgfplotsset{compat=1.14}

\usetikzlibrary{arrows,shapes}
\usetikzlibrary{shapes.geometric}
\newcolumntype{.}{D{.}{.}{-1}}
\newcolumntype{.}{D{.}{.}{-1}}
\graphicspath{{./Tables/}}

\begin{document}
\sloppy
\font\myfont=cmr12 at 9pt
\newgeometry{margin=0.75in}
\title{Robust Tests for Convergence Clubs\thanks{%
{\myfont We are grateful to Andrew Harvey, Hashem Pesaran, Jonathan Temple, Mike
Wickens, Gernot Dopplehoffer, Vasco Carvalho for their comments on the work.}}}
\author{Luisa Corrado\thanks{%
{\myfont Luisa Corrado gratefully acknowledges the Marie-Curie
Intra European fellowship 039326. luisa.corrado@uniroma2.it}}\smallskip \\
%EndAName
University of Rome Tor Vergata and University of Cambridge \and Thanasis
Stengos \\
%EndAName
University of Guelph \and Melvyn Weeks\smallskip \\
%EndAName
Faculty of Economics and Clare College, University of Cambridge \and M. Ege
Yazgan \\
%EndAName
Istanbul Bilgi University}
\maketitle

\begin{abstract}
\begin{comment}The extent \todo[fancyline]{\tiny All: Mention the MC results? The abstract
needs to been rewritten. Remove from "The exent to which".....to "regional
development policies". Start from "We consider tests....} to which countries
and or regions are similar across one or more dimensions is a question that
has long been of interest to economists and policymakers. The correct
identification of the extent of convergence within a regional economy is
paramount given that policy usually tries to achieve convergence by reducing
the gap between the richest and the poorest regions. Within the European
Union the ECB targets a \emph{single} Euro Area inflation rate, and in this
respect the degree to which there exists convergence in regional per capita
incomes and output is of critical relevance to European regional development
policies.
\end{comment}
\singlespacing

{\small In many applications common in testing for convergence the number of
cross-sectional units is large and the number of time periods are few. In
these situations asymptotic tests based on an omnibus null hypothesis are
characterised by a number of problems. In this paper we propose a multiple
pairwise comparisons method based on an a recursive bootstrap to test for
convergence with no prior information on the composition of convergence
clubs. Monte Carlo simulations suggest that our bootstrap-based test
performs well to correctly identify convergence clubs when compared with
other similar tests that rely on asymptotic arguments. Across a potentially
large number of regions, using both cross-country and regional data for the
European Union we find that the size distortion which afflicts standard
tests and results in a bias towards finding less convergence, is ameliorated
when we utilise our bootstrap test.}

\bigskip


\medskip

\end{abstract}

\sloppy
\font\myfont=cmr12 at 11pt \newgeometry{margin=0.75in}

%\newpage

%\tableofcontents

\newpage

\singlespacing

\pagebreak

\doublespacing
\restoregeometry
%\renewcommand{\baselinestretch}{1.5}
%\setstretch{2}

\pagebreak

\section{Introduction}

The extent to which countries and or regions are similar across one or more
dimensions is a question that has long been of interest to economists and
policymakers. Within the European Union the ECB targets a single Euro Area
inflation rate, and in this respect the degree to which there exists
convergence in regional per capita incomes and output is of critical
relevance to European regional development policies (Boldrin and Canova
(2001)). Moreover, one of the core components of the European cohesion
policy has been to reduce the disparities between income levels of different
regions and in particular the backwardness of the least favoured regions;
this objective has, in general, been manifest as the promotion of
convergence between \textsc{eu} regions.\footnote{%
See Article 158 of the Treaty establishing the European Community.} In this
context it is evident that the correct detection of the extent of
convergence within a regional economy is paramount given that policy usually
tries to achieve regional convergence by reducing the gap between the
richest and the poorest regions. In this respect a test of convergence which
exhibits bias, for example being oversized in small samples, will mislead,
and in this instance imply less convergence suggesting the need for more
policy initiatives than may actually be required.

Economists have conceptualised the notion of similarity using formal
definitions of convergence based upon growth theory. Standard neoclassical
growth models (Solow (1956) and Swann (1956)) founded upon the key tenets of
diminishing returns to capital and labour and perfect diffusion of
technological change, dictate that countries will converge to the same level
of per capita income (output) in the long run, independent of initial
conditions. The New Growth theory (see, for example, Romer (1986); Lucas
(1988); Grossman and Helpman (1994); Barro and Sala-i-Martin (1997)) allows
for increasing returns to accumulable factors such as human capital in order
to determine the (endogenous) long-run growth rate.\footnote{%
Other variants of the New Growth theories predict the emergence of multiple
locally-stable steady-state equilibria instead of the unique globally-stable
equilibrium of the neoclassical growth model as a result of differences in
human and physical capital per worker across countries (Basu and Weil
(1998)), their state of financial development (Acemoglu and Zilibotti
(1997)) or other externalities caused by complementarity in innovation
(Ciccone and Matsuyama (1996)).} The emergence over the past decade of New
Economic Geography\footnote{%
In the `new economic geography' models the sources of increasing returns are
associated with Marshallian-type external localisation economies (such as
access to specialised local labour inputs, local market access and size
effects, local knowledge spillovers, and the like). These models provide a
rich set of possible long run regional growth patterns that depend, among
other things, on the relative importance of transport costs and localisation
economies (Fujita, et al. 1999; Fujita and Thisse 2002).} models of
industrial location and agglomeration, has resulted in the identification of
other forces which generate increasing returns, two notable examples being
the relationship between location and transportation costs (Louveaux \textit{%
et al}., 1982) and the effect of regional externalities (Cheshire and Hay
(1989)).

To the extent that the process of growth is different across regions in the
sense that there are different long-run steady-states, the standard
neoclassical growth model is not valid. In this context traditional
approaches to test for convergence are hard to justify, difficult to
interpret, and often difficult to implement. For example, a rejection of the
omnibus null of convergence across a groups of regions provides increasingly
less information as the number of regions increases and where prior
knowledge over both the number and composition of convergence clubs is
minimal. Moreover, the justification of constructing such a large
intersection null hypothesis is often questionable at the outset. Faced with
the emergence of larger panels, with an attendant increase in
cross-sectional heterogeneity, there has been a number of significant
developments in testing. For example, the use of a heterogeneous alternative
hypothesis partially alleviates the problem of testing over a large group of
potentially heterogenous regions (see, for example, Im et al., 2003).

In a further progression away from the testing of general omnibus
hypotheses, Pesaran (2007) conducts pairwise tests for region pairs, with
inference focussed on the proportion of output gaps that are stationary. One
drawback of this approach is that limited inference can be made as to the
significance of individual gaps, or indeed whether a group of output
comparisons form a convergence club. An approach which allows for an
endogenous determination of the number of clubs using a sequence of pairwise
stationarity tests has been developed by Hobijn and Franses (2000). In
extending this approach Corrado, Martin and Weeks (2005) developed a testing
strategy that facilitates both the endogenous identification of the number
and composition of regional clusters (or `clubs'), and the interpretation of
the clubs by comparing observed clusters with a number of hypothesized
regional groupings based on different theories of regional growth. However,
given that the time series are relatively short, there are potential
problems in basing inference on asymptotic results for stationarity tests.
Reliance on large $T$ asymptotics is likely to impart a size distortion,
biasing the results towards finding less convergence than actually exists.

%examined the evidence for regional convergence across the European
%Union.

To circumvent this problem we propose in this paper a recursive bootstrap
test for stationarity which is designed to detect multiple convergence clubs
without prespecification of group membership. Monte Carlo simulations
suggest that the proposed bootstrap based method performs quite well in
identifying club membership when compared with the Hobijn and Franses (2000)
approach that is based on asymptotic arguments. We implement our bootstrap
recursive test of convergence using the original cross-country dataset used
by Hobijn and Franses (2000) and the European regional data used in Corrado,
Martin and Weeks (2005). We then compare the asymptotic and bootstrap
generated cluster outcomes. Our results show that by resolving the size
distortion which afflicts the asymptotic test we find considerably more
evidence of convergence in both the applications considered.

The paper is structured as follows. Section two reviews existing tests for
convergence clubs, and in section three we present the bootstrap version of
the test. In section four we propose a Monte Carlo experiment to compare the
properties of the asymptotic and bootstrap tests. Section five describes the
data and applies the proposed tests to two real word datasets. In section
six we discuss our findings and conclusions.

\section{Tests for Convergence}

%[MELVYN: I AM UNSURE WHETHER WE SHOULD INSERT EQ 1 AND 2 Answer: Eq 1 and 2 are different definitions of convergence, I think they are fine as they are]

In this section we briefly discuss a number of significant developments in
tests designed to detect convergence and identify clubs which are able to
address a number of questions such as whether a particular pair of countries
have converged, or whether a group of regions or countries form a
convergence club. We briefly discuss the different approaches to detect
convergence tracking a gradual progression away from multivariate time
series and panel data tests based on an omnibus null, towards sequential
tests and tests that are founded upon multiple pairwise comparisons.%
\footnote{%
Corrado and Weeks (2011) provide a more detailed overview.} Our focus here
is to identify endogenously clubs using multivariate tests for stationarity.
However, given that the time series are relatively short, we show that there
are potential problems in basing inference on asymptotic results for
stationarity tests. To circumvent this problem we bootstrap the stationarity
test and assess the effect of the size distortion on the cluster outcomes
using two different applications based on country and regional level data.

The use of \textit{multivariate time-series} to test for convergence was
initiated by the seminal papers of Bernard and Durlauf (1995, 1996). Given a
set $\digamma$ of $N$ economies, a multi country definition of relative
convergence asks whether the long-run forecast of all output differences
with respect to a benchmark economy, (denoted with the subscript $1$) tend
to a country-specific constant as the forecasting horizon tends to infinity.%
\footnote{%
A necessary condition for regions $i$ and $j$ to converge, either absolutely
or relatively, is that the two series must be cointegrated with
cointegrating vector $[1,-1]$. However, if output difference are trend
stationary, this implies that the two series are co-trended as well as
cointegrated. Hence a stronger condition for convergence is that output
differences cannot contain unit roots or time trends (Pesaran (2007)).} We
may then write
\begin{equation}
\lim_{s\rightarrow \infty }E(y_{(i1),t+s}\mid I_{t})=\mu _{1i}\text{ \ \ \ \
\ \ \ \ }\forall i\neq 1,\text{ \ \ }
\end{equation}%
where $y_{(i1),t+s}=y_{it+s}-y_{1t+s}$ and $\mu _{1i}$ is a finite constant.%
\footnote{%
We consider this as a more reasonable definition of convergence in the sense
that it allows the process of convergence to stop within a neighborhood of
zero mean stationarity (absolute convergence) and is consistent with the
existence of increasing costs of convergence.} There exist a number of
problems with multivariate time series tests. First, the testing procedure
is sensitive to the choice of the benchmark country. Second, in keeping with
the problems of omnibus tests, in the event of rejecting the non-convergence
null we have no information as to which series are $I(0)$ and $I(1)$, nor
the composition of any convergence groups. Third, given the system
properties of the test, a dimensionality constraint means that it can handle
only a small number of economies simultaneously.

\emph{Panel} unit root procedures have also been adopted to test for
convergence by considering the stationary properties of output deviations
with respect to a benchmark economy (Fleissig and Strauss, 2001; Evans,
1998; Carlino and Mills, 1993). First, the so called `first-generation'
panel unit-root tests,\footnote{%
See, for example, Maddala and Wu, 1999; Im et al., 2003; Levin, Lin and Chu,
2002.} maintain that errors are independent across cross-sectional units
which imparts a size distortion. To overcome this problem a `second
generation' of panel unit root tests have been developed which allows for
different forms of cross-sectional dependence.\footnote{%
For example, Taylor and Sarno (1998) adopt a multivariate approach and
estimate a system of $N-1$\ ADF equations using Feasible GLS to account for
contemporaneous correlations among the disturbances. Other notable example
of second generation of panel unit root tests with cross-sectional
dependence include Pesaran (2007) and Moon and Perron (2007).} However, as
pointed out by Breitung and Pesaran (2008), panel data unit root tests poses
similar problems in that as $N$ becomes large the likelihood of rejecting
the omnibus null increases with no information on the exact form of the
rejection.

The problem of identifying the mix of $I(0)$ and $I(1)$ series whilst still
utilising the attendant power from a panel by exploiting coefficient
homogeneity under the null, has been addressed by the \textit{sequential}
test proposed by Kapetanios (2003).\footnote{%
See also Flores et al. (1999) and Breuer et al. (1999).} Specifically,
Kapetanios employs a sequence of unit root tests of panels of decreasing
size to separate stationary and nonstationary series,\footnote{%
This method is referred to as the Sequential Panel Selection Method (SPSM).}
facilitating an endogenous identification of the number and identity of
stationary series. Although a positive development there are a number of
limitations. Critically the utility of this approach depends on the use of a
panel framework to add power in a situation where most series are stationary
but very persistent. In addition, the method only permits the classification
of the $N$ series into two groups whereas there may be many more groups. As
a consequence it is not possible to address a number of questions that may
be of interest: such as whether a particular pair of countries have
converged, or whether a group of regions or countries form a convergence
club.

%\fxnote{How is the multicountry defn different than eqn (1)? Answer: The presence of a benchmark in eq (1) }
When applied to output deviations, an additional problem with the Kapetanois
test is that the testing procedure is still sensitive to the choice of the
benchmark country. One approach which avoids the pitfalls of the choice of a
benchmark country and, more generally, the dimensionality problem that
afflicts the application of omnibus tests, is to conduct separate tests of
either stationarity and/or nonstationarity. By considering a particular
multi-country definition of convergence, Pesaran (2007) adopts a \textit{%
pairwise} approach to test for unit-roots and stationarity properties of all
$N(N-1)/2$ possible output pairs $\left\{ y_{it}-y_{jt}\right\} $. The
definition of convergence that is adopted is that the $N$ countries converge
if
\begin{equation}
\Pr (\cap _{i=1,...,N,j=i+1,...,N}|y_{it+s}-y_{jt+s}|<c|I_{t})>\pi ,
\end{equation}%
%
%\todo[fancyline]{\tiny \emph{All}: Check notation for eqn (2)}
for all horizons, $s=1,...,\infty $, and $c$ a positive constant. $\pi \geq 0
$ denotes a tolerance probability which denotes the proportion that one
would expect to converge by chance.

In testing the significance of the \emph{proportion} of output gaps that
indicate convergence, the dimensionality constraint that affects the
application of system-wide multivariate tests of stationarity is
circumvented. However, although the pairwise tests of convergence proposed
by Pesaran (2007) is less restrictive than the omnibus tests proposed by
Bernard and Durlauf (1995), the subsequent inference is limited in that it
does not allow inference on which pairs of regions have converged, or the
number and composition of convergence clubs. Below we examine various
testing strategies for club convergence and in particular the \textit{%
sequential testing procedure} proposed by Hobijn and Franses (2000).

\subsection{Sequential Pairwise Tests}

Despite the use of multivariate time series and panel data methodologies to
test for convergence, there has been relatively few attempts to utilize this
approach to systematically identify convergence clubs (see Durlauf \textit{%
et al.} 2005). The existing early methods were generally focused on the
convergence of various \textit{a-priori }defined homogeneous country groups
which were assumed to share the same initial conditions. Baumol (1986) for
example grouped countries with respect to political regimes (OECD
membership, command economies and middle income countries), Chatterji (1992)
allowed for clustering based on initial income per capita levels and tested
convergence cross-sectionally, while Durlauf and Johnson (1995) grouped
countries using a regression tree method based on different variables such
as initial income levels and literacy rates that determined the different
"nodes" of the regression tree. Similarly, Tan (2009), utilizes a regression
tree approach which again utilises exogenous information in the form of
conditioning variables. An alternative approach to the cross-sectional
notion of $\beta -$ convergence was introduced by Bernard and Durlauf (1995,
1996) based on a time series framework that makes use of unit root and
cointegration analysis (see Durlauf \textit{et al.} (2005) for a
comprehensive literature review for convergence hypothesis).

There are many other studies where the identification of convergence clubs
has been achieved exogenously through testing in conjunction with a
pre-classification of clubs using parametric techniques.\footnote{%
See Quah (1997) for an example of non-parametric techniques to locate
convergence clubs.} For example, Weeks and Yao (2003) adopt this approach
when assessing the degree of convergence across coastal and interior
provinces in China over the period 1953-1997. As Maasoumi and Wang (2008)
note, the principal problem with pre-classification is that as the number of
regions increases such a strategy is not robust to the existence of other
convergence clubs within each sub-group. A different approach is advanced
using the notion of $\sigma -$convergence by Phillips and Sul (2007) who
developed an algorithm based on a log-$t$ regression approach that clusters
countries with a common unobserved factor in their variance.

In general it is straightforward to test whether two regions form a single
group. We could simply construct a single output (income) deviation and test
for stationarity or a unit root. In the context of differentiating between
the stationarity properties of multiple series (or output deviations), the
contribution by Kapetanios (2003) has provided new techniques that utilise
the power of an omnibus null in conjunction with a sequential test that
allows greater inference under the alternative hypothesis. However, for a
large set of regions, $\digamma $, locating partitions that are consistent
with a particular configuration of convergence clubs generates difficulties
given that the number of combinations is large and related, that we have
little prior information.\footnote{%
Harvey and Bernstein (2003), utilize non-parametric panel-data methods
focussing on the evolution of temporal level contrasts for pairs of
economies, identifying the number and composition of clusters. Beylunio\u{g}%
lu et al (2018) also propose a method based on the maximum clique method
from graph theory that relies on Augmented Dickey Fuller (\textsc{adf}) unit
root testing. However, as it will be explained below, the maximum clique
method is a ``top down'' method that leads to a different definition of
clubs and is not directly comparable to the ``bottom up'' Hobijn and
Franses' approach.} Hobijn and Franses (2000) propose an empirical procedure
that endogenously locates groups of similar countries (convergence clubs)
utilising a sequence of stationarity tests. Cluster or club convergence in
this context implies that regional per capita income differences between the
members of a given cluster converge to zero (in the case of absolute
convergence) or to some finite, cluster specific non-zero constant (in the
case of relative convergence). Below we illustrate the method.

The Hobijn and Franses (2000) test represents a multivariate extension of
the Kwiatkowski \textit{et al.}(1992) test (hereafter KPSS test). We
introduce the test by first denoting $\mathbf{y}_{t}=\{y_{it}\}$ as the $%
N\times 1$\ vector of log per capita income and write $\boldsymbol{y}_{t}$
as
\begin{equation}
\boldsymbol{y}_{t}=\boldsymbol{\alpha }+\boldsymbol{\beta }t+\mathbf{D}%
\sum_{s=1}^{t}\mathbf{v}_{s}+\bm{\varepsilon }_{t},  \label{compoH-F}
\end{equation}%
where $\boldsymbol{\alpha }=\mathbf{\{}\alpha _{i}\}$ is a $N\times 1$
vector of constants, $\boldsymbol{\beta }=\mathbf{\{}\beta _{i}\}$ is a $%
N\times 1$\ vector of coefficients for the deterministic trend $t$, and $%
\mathbf{v}_{s}=\{v_{l,s}\}$, $l=1,...,m$ represents a $m\times 1$ vector of
first differences of the $m$ stochastic trends in $\mathbf{y}_{t}$, $m\in
(0,...,N) $. $\mathbf{D}=\{D_{i,l}\}$ denotes a $N\times m$ matrix with
element $D_{i,l}$ denoting the parameter for the $l^{th}$ stochastic trend. $%
\boldsymbol{\varepsilon }_{t}=\{\varepsilon _{i,t}\}$ is a $N\times 1$
vector of stochastic components.

In considering the difference in log per capita income for regions $i$ and $%
j $ we write
%\fxnote{Why the sum $\sum_{l=1}^{m}$ in (4) - does this make sense when just considering regions $i,j$? YES SINCE WE DEFINE LATER $D_{(ij),l}= D_{il}-D_{jl}=0\text{ \ }\forall l=1,...,m$ FROM TABLE I HF HENCE SUM OVER $l$ IS CORRECT}
\begin{equation}
y_{(ij),t}=\alpha _{(ij)}+\beta _{(ij)}t+\sum_{l=1}^{m}D_{(ij),l}\bigg(%
\sum_{s=1}^{t}v_{l,s}\bigg)+\varepsilon _{(ij),t},  \label{pair}
\end{equation}%
where (\ref{pair}) admits two different convergence concepts: absolute and
relative. The restrictions implied by the null of relative convergence are $%
\beta _{(ij)}=0\text{ \ }\forall i\neq j\in \digamma $ and $%
D_{(ij),l}=D_{il}-D_{jl}=0\text{ \ }\forall l=1,...,m$, with the latter
restriction indicating that the stochastic trends in log per capita income
are cointegrated with cointegrating vector $[1-1]$. The additional parameter
restrictions for the null hypothesis of absolute convergence are that $%
\alpha _{(ij)}=0\text{ \ }\forall i\neq j\in \digamma $. Both asymptotic
absolute and relative convergence imply that the cross sectional variance of
log per capita income converges to a finite level.

Denoting the partial sum process $S_{t}=\sum\limits_{s=1}^{t}y_{(ij),s}$,
the test statistic for zero mean stationarity is given by\footnote{%
For the stationary null, Hobijn and Franses (2000) utilise the Kwiatkowski
\textit{et al.} (1992) test. The KPSS test is operationalised by regressing
the pairwise difference in per capita income $y_{(ij),t}$ against an
intercept and a time trend giving residuals
\begin{equation}
\widehat{\varepsilon }_{(ij),t}=y_{(ij),t}-\widehat{\alpha }_{(ij)}-\widehat{%
\beta }_{(ij)}t.
\end{equation}%
and%
\begin{equation}
\widehat{\sigma }^{2}=\frac{1}{T}\sum\limits_{t=1}^{T}\widehat{\varepsilon }%
_{(ij),t}^{2}+2\frac{1}{T}\sum\limits_{k=1}^{L}\omega
(k,L)\sum\limits_{t=k+1}^{T}\widehat{\varepsilon }_{(ij),t}\widehat{%
\varepsilon }_{(ij),t-k},  \notag
\end{equation}%
represents the consistent Newey-West estimator of the long-run variance. $%
\omega (k,L)=1-k/(1+L)$, $k=1,...,L$ is the Bartlett kernel, where $L$
denotes the bandwidth.}
\begin{equation}
\widehat{\tau }_{0}=T^{-2}\sum\limits_{t=1}^{T}S_{t}^{^{\prime }}[\widehat{%
\sigma }^{2}]^{-1}S_{t}.  \label{zero}
\end{equation}%
Denoting $h_{t}=y_{(ij),t}-\frac{1}{T}\sum\limits_{t=1}^{T}y_{(ij),t}$ and
the partial sum process as $\bar{S}_{t}=\sum\limits_{s=1}^{t}h_{s}$ the test
statistic for level stationarity is given by

\begin{equation}
\widehat{\tau }_{\mu }=T^{-2}\sum \limits_{t=1}^{T}\bar{S}_{t}^{^{\prime }}[%
\widehat{\sigma }^{2}]^{-1}\bar{S}_{t}.  \label{level}
\end{equation}

Examining (\ref{pair}), we note that in the case of two regions, and
focussing on a test of relative convergence with restrictions $D_{(ij),l}=0%
\text{ \ }\forall l=1,...,m$ and $\beta _{(ij)}=0$, it is obviously
straightforward to test whether two regions form part of a single group.
However, for a large number of regions locating the partitions over $%
\digamma $ that are consistent with a particular configuration of
convergence clubs is infeasible both because the number of combinations is
large and related, that we have little prior information on the form of $%
\mathbf{D}$ and the likely combination of zeros restrictions over the
differences $\beta _{(ij)}$ and $\alpha _{(ij)}$. The alternative testing
strategy proposed by Hobijn and Franses (2000) forms groups from the bottom
up using a clustering methodology to determine, endogenously, the most
likely combination of restrictions, and as a consequence, the most likely
set of convergence clubs. The cluster algorithm is based on the hierarchical
farthest neighbour method due to Murtagh (1985). We illustrate the
sequential test using the set of regions $\digamma =\{1,2,3,4\}$.

\begin{enumerate}
\item[ (i) ] We first initialise singleton clusters $K(i)$ for each region $%
i=1,...,4$. The null hypothesis of level stationarity is tested for all $%
N(N-1)/2=6$ region pairs. We collect $p$-values in the vector $\widehat{%
\mathbf{p}}_{s=1}=\{p_{(ij)}\}$, where $p_{(ij)}=\Pr (\widehat{\tau }%
_{(ij),\mu }<c_{(ij)}|I_{t})$, $\widehat{\tau }_{(ij),\mu }$ denotes the
test statistic and $c_{(ij)}$ the critical value. $s=1$ denotes the first
iteration.

Clusters are formed on the basis of the max $p$-value in $\widehat{\mathbf{p}%
}_{s=1}$, indicating the pair of regions which are most likely to converge.
If, for example, $p_{(1,2)}=\{\underset{i,j\in \digamma }{\max }\{\widehat{%
\mathbf{p}}_{s=1}\}>p_{\mathtt{min}}\}$ then regions 1,2 are the first pair
of regions to form a club.\footnote{%
The choice of $p_{\min }$\ has a direct effect on the cluster size. Since
the stationarity test is known to be oversized in small samples, this bias
will generate inference towards finding less convergence.} We denote the
first cluster as $K(1^{^{\prime }})=\{1,2\}$ and discard the singleton
cluster 2, which is now part of the two-region cluster $K(1^{^{\prime }})$.

\item[ (ii)] In the second iteration ($s=2$) we define the set of regions as
$\digamma ^{^{\prime }}=(1^{^{\prime }},3,4)$. We form pairwise output
differences between the $N-2$ remaining singleton clusters and the
two-region cluster $K(1^{\prime })$.
Once again we collect the $p$-values in the vector $\widehat{%
\mathbf{p}}_{s=2}$. Letting $p_{(r,v)}=\{\underset{i,j\in \digamma }{\max }\{%
\widehat{\mathbf{p}}_{s=2}\}>p_{\mathtt{min}}\}$, then if, for example, $%
p_{(r,v)}=p_{(1^{\prime }3)}$, the singleton cluster $K(3)$ joins cluster $%
K(1^{\prime })$ forming a three-region cluster $K(1^{^{\prime \prime
}})=\{1,2,3\}.$

\item[ (iv)] In this example we find a three-region cluster $K(1^{^{\prime
\prime }})$ and a singleton cluster $K(4),$ so the procedure stops.

%More
%generally, the testing procedure then iterates through the ordered $p$%
%-values, confronting the max with the minimum p-value, where at
%each iteration the max is falling.
\end{enumerate}

The principle difference between this sequential testing strategy and the
SPSM approach of Kapetanios is that the SPSM test is designed to
endogenously classify stationary and nonstationary series. This is achieved
by sequentially reducing the size of the omnibus null by removing series
with the most evidence against the unit root null, classifying these series
as stationary. The stopping point is when the unit root null does not
reject, such that all the remaining regions are declared nonstationary. In
contrast the Hobijn and Franses method seeks to endogenously allocate $N$
series to $J\leq N$ convergence clubs. This is achieved by only classifying
regions that provide, at each recursion and conditional on exceeding $p_{%
\mathtt{min}}$, the most evidence for convergence.

Although the sequential multivariate stationarity test is consistent in that
for large $T$ the tests will reveal the true underlying convergence clubs,
the principle shortcoming is that the test statistic is known to be
oversized in small samples (Caner and Kilian, 2001). When testing for
convergence using yearly data $T$ is likely to be small, and as a result
inference is likely to be biased in the direction of finding less
convergence. Similar size distortions also emerge when the series are
stationary but highly persistent: in this case the partial sum of residuals
which are used to derive the KPSS test resemble those under the alternative
in the limit. Below we outline a bootstrap approach which circumvents the
pitfalls of inference based upon asymptotic arguments since it is able to
generate independent bootstrap resamples using a parametric model which is
conditional on the sample size and the dependence structure of the dataset.
In section 5 we utilise this test to investigate the extent of convergence
in two different applications: (i) the cross-country dataset originally
adopted by Hobijn and Franses (2000); (ii) the European regional dataset
used by Corrado, Martin and Weeks (2005).

\section{A Bootstrap Test}

To derive the parametric model with which to create independent bootstrap
samples under the stationarity null, following Kuo and Tsong (2005) and
Leybourne and McCabe (1994), we exploit the equivalence in second order
moments between an unobserved component model and a parametric ARIMA model
(Harvey (1989)) for the differenced data. In demonstrating this equivalence
we note that (\ref{pair}) may be rewritten in structural form as a function
of a deterministic component ($\alpha _{(ij)}+\beta _{(ij)}t$), a random
walk ($r_{t}$) and a stationary error ($\varepsilon _{(ij),t}$):
\begin{eqnarray}
y_{(ij),t} &=&\alpha _{(ij)}+\beta
_{(ij)}t+\sum_{l=1}^{m}D_{(ij),l}r_{l,t}+\varepsilon _{(ij),t}  \label{compo}
\\
r_{l,t} &=&r_{l,t-1}+v_{l,t},  \label{compo1}
\end{eqnarray}%
where $r_{l,t}=$ $\sum_{s=1}^{t}v_{l,s}$ represents the $l$-th stochastic
trend for regions $i$ and $j$ with $r_{0}$, the fixed initial value, set to
zero. We also assume that $\varepsilon _{(ij),t}$ is a stationary error
process $\varepsilon _{(ij),t}=\sum_{s=0}^{\infty }\psi
_{(ij),s}u_{(ij),t-s}=\Psi (L)u_{(ij),t}$ where $\psi _{(ij),0}=1$, $%
u_{(ij),t}\sim i.i.d(0,\sigma _{u_{(ij)}}^{2})$ and $\Psi
(L)=1+\sum_{s=1}^{\infty }\psi _{(ij),s}L^{s}$.\footnote{$\varepsilon
_{(ij),t}$ is assumed to be invertible $\sum_{s=0}^{\infty }s\left\vert \psi
_{(ij),s}\right\vert <\infty $.} Under these assumptions $\left\{
\varepsilon _{(ij),t}\right\} $ has an infinite order autoregressive
representation %\fxnote{Do we need the $(ij)$ notation for $\psi
%_{(ij),s}$ see page 5 Kuo and Tsong? YES THE PARAMETER FOR EACH LAG REFERS TO A SPECIFIC PAIR OF REGIONS}
\begin{equation}
\varepsilon _{(ij),t}=\sum_{s=1}^{\infty }\lambda _{(ij),s}\varepsilon
_{(ij),t-s}+u_{(ij),t},
\end{equation}%
where $\Lambda (L)=\Psi (L)^{-1}=1+\sum_{s=1}^{\infty }\lambda
_{(ij),s}L^{s} $. Given (\ref{compo}), since $\left\{ \varepsilon
_{(ij),t}\right\} $ is a stationary process, the necessary condition for
convergence of regions $i,j$ is that the variance of the random walk error ($%
\sigma _{v}^{2}$) is zero. Focussing on a test for relative convergence,
below we describe the nature of the recursive multivariate stationarity test
using critical values generated from the empirical distribution of the test
statistic constructed using bootstrap sampling.

In generating a bootstrap test for relative convergence we focus on relative
convergence where $\beta _{(ij)}=0$, which rules out the presence of a
deterministic trend.\footnote{%
For the test of absolute convergence the restrictions are $\beta
_{(ij)}=\alpha _{(ij)}=0.$} The idea is to estimate the null finite sample
distribution of the KPSS test statistics by exploiting the equivalence
between the unobservable component model and the parametric ARIMA model.
Harvey (1989) demonstrates that the components from the structural model (%
\ref{compo}) can be combined to give a reduced form ARIMA(0,1,1) model. In
particular, assuming independence of $\varepsilon _{(ij),t}$ and $v_{t}$, (%
\ref{compo}) becomes a local component model which, after time differencing,
can be expressed as the MA model $\Delta y_{(ij),t}=(1-\theta L)\eta
_{(ij),t}$ where $\eta _{(ij),t}\sim i.i.d(0,\sigma _{\eta _{(ij)}}^{2})$
and $\sigma _{\eta _{(ij)}}^{2}=\sigma _{\varepsilon _{(ij)}}^{2}/\theta .$

The reduced form parameter $\theta $ is derived by equating the
autocovariances of first differences at lag one in the structural and
reduced forms. This gives the following relationship between the parameters
of the components model (\ref{compo}) and the ARIMA(0,1,1) model:
\begin{equation}
\theta =\frac{1}{2}\left \{ \frac{\sigma _{v}^{2}}{\sigma _{\varepsilon
_{(ij)}}^{2}}+2-\left( \frac{\sigma _{v}^{2}}{\sigma _{\varepsilon
_{(ij)}}^{2}}+4\frac{\sigma _{v}^{2}}{\sigma _{\varepsilon _{(ij)}}^{2}}%
\right) ^{1/2}\right \}.
\end{equation}%
where $q=\frac{\sigma _{v}^{2}}{\sigma _{\varepsilon _{(ij)}}^{2}}$ is the
signal to noise ratio. Under the stationarity null, namely that regions $i,j$
are converging, the variance of the random walk component ($\sigma _{v}^{2}$%
) is zero, which in turn implies that $\theta =1$ in the ARIMA
representation. Therefore by imposing a moving average unit root in the
ARIMA representation one can use the parametric model for sampling instead
of the "unobservable" component model.

Our bootstrap sampling scheme is based on the following procedure. First,
for each region pair $i,j$ and contemporaneous difference $%
y_{(ij),t}=y_{i,t}-y_{j,t}$, we fit an ARMA($p,1$) model to the differenced
series $\Delta y_{(ij),t}=y_{(ij),t}-y_{(ij),t-1}$, namely
\begin{equation}
\Delta y_{(ij),t}=\sum_{k=1}^{p}\phi _{(ij),k}\Delta y_{(ij),t-k}+\eta
_{(ij),t}-\theta \eta _{(ij),t-1},  \label{ARrep}
\end{equation}%
The MA component in (\ref{ARrep}) follows from the reparametrisation of the
structural component model to reproduce the stationarity properties of the
data in the ARMA representation. The AR(p) component\footnote{$p$ denotes
optimal lag length, chosen using the AIC criterion.} represents an
approximation to the assumed infinite-order moving average errors to capture
the dependence structure in the data. By imposing a moving average unit root
in the sampling procedure we can then construct the bootstrap distribution
of the test statistic for level stationarity defined in (\ref{level}).

The accuracy of the bootstrap test relative to the asymptotic approximations
hinges on the bootstrap sample being drawn independently. Given the presence
of a known dependence structure, in this case a stationary ARMA($p,1$)
model, we utilise the Recursive Bootstrap.\footnote{%
See Horowitz (2001) on the merits of the recursive bootstrap for linear
models, and Maddala and Li (1997) and Efron and Tibshirani (1986) for
specific examples.} To achieve independent re-sampling from (\ref{ARrep}) we
estimate $\hat{\phi}_{(ij),k}$ and $\hat{\eta}_{(ij),t}$, and we draw a
bootstrap sample $\{\bar{\eta}_{(ij),t}^{r}\}_{t=1}^{T}$ from the
distribution of centered\footnote{%
Centering the residuals reduces the downward bias of autoregression
coefficients in small samples (Horowitz, 2001).} residuals $\{\bar{\eta}%
_{(ij),t}\}_{t=1}^{T}$, where $\bar{\eta}_{(ij),t}=\widehat{\eta }_{(ij),t}-%
\frac{1}{T-1}\sum_{t=2}^{T}\widehat{\eta }_{(ij),t}$.

Given the bootstrapped residuals, $\{\bar{\eta}_{(ij),t}^{r}\}_{t=1}^{T}$ ,
the $r^{th}$ bootstrap sample for the data $\{\Delta
y_{(ij),t}^{r}\}_{t=1}^{T}$ is generated based on the recursive relation%
\footnote{%
Initial values, $\Delta
y_{(ij),t-1}^{r}=y_{(ij),t-1}^{r}-y_{(ij),t-2}^{r}=...=\Delta
y_{(ij),t-p}^{r}=y_{(ij),t-p}^{r}-y_{(ij),t-p-1}^{r}$ are set to zero.}
\begin{equation}
\Delta y_{(ij),t}^{r}=\sum_{k=1}^{p}\widehat{\phi }_{(ij),k}\Delta
y_{(ij),t-k}^{r}+\bar{\eta}_{(ij),t}^{r}-\bar{\eta}_{(ij),t-1}^{r}.
\label{rec}
\end{equation}%
We then recover the level of the series (where the level denotes the
contemporaneous regional difference) directly from (\ref{rec})
\begin{equation}
y_{(ij),t}^{r}=y_{(ij),t-1}^{r}+\sum_{k=1}^{p}\widehat{\phi }_{(ij),k}\Delta
y_{(ij),t-k}^{r}+\eta _{(ij),t}^{r}-\eta _{(ij),t-1}^{r}.
\end{equation}%
Defining $h_{t}^{r}=y_{(ij),t}^{r}-\frac{1}{T}\sum%
\limits_{t=1}^{T}y_{(ij),t}^{r}$, then for $r^{th}$ bootstrap sample, and
the $i,j$ region pair, a test statistic for relative convergence, $\widehat{%
\tau }_{(ij),\mu }^{r}$, is given by
\begin{equation}
\widehat{\tau }_{(ij),\mu }^{r}=T^{-2}\sum\limits_{t=1}^{T}\bar{S}%
_{t}^{^{r,\prime }}[\widehat{\sigma }^{r,2}]^{-1}\bar{S}_{t}^{r},
\end{equation}%
where $\bar{S}_{t}^{r}=\sum\limits_{s=1}^{t}h_{s}^{r}$. For each region pair
we draw $R$ bootstrap samples and construct the empirical distribution of
the test statistic under the null, which we denote $\tau _{(ij),\mu }^{B}$.
Bootstrap critical values $C_{(ij),\mu }^{B}$ can then be recovered at the
required significance levels and we can implement the algorithm described in
section 3 utilising a vector of bootstrapped empirical p-values, $\mathbf{%
\hat{p}}^{B}$.

\section{A Monte Carlo Study}

In this section we compare the performance of the bootstrap version of the
\textsc{kpss} test, \textsc{cw} henceforth, with the original \textsc{hf}
test based on the asymptotic version of the multivariate \textsc{kpss}
testing procedure. To the best of our knowledge there is no other comparable
Monte Carlo study in the literature that evaluates clustering methods in the
same context as we do here.

\textsc{hf} is a method that relies on a ``bottom up'' algorithm that
clusters groups one by one. To determine whether a set of countries is
convergent, \textsc{hf} applies a multivariate stationarity test to panels
comprised of consecutive pairwise difference series set elements and
confirms convergence if the null hypothesis of stationarity of the panel is
not rejected using the \textsc{kpss} test. For example, if we want to test
the convergence of countries 1,2,3 and 4, a panel consisting of $%
y^{12},y^{13},$ $y^{23},y^{14},y^{24}$ and $y^{34}$ is subjected to the
\textsc{kpss} test, where for example $y^{12}$ ,$y^{23},$ and $y^{34}$
denote the difference in log per capita between countries 1 and 2, 2 and 3
and 3 and 4 respectively. If stationarity cannot be rejected the panel is
then augmented with series other than 1, 2, 3 and 4, each added separately.
%Is this description of the H-F method consistent with our description in section 2.1: Answer: The HF is a bottoms up method and it uses all pairs but in a consecuitve manner, augmenting the group until convergence is achieved. for the given group of countries.
If then for each of these additional panels the stationarity null is
rejected, then these four countries are said to be convergent.

\subsection{Monte Carlo Structure}

In this subsection, we will discuss the data generating processes that is
used in our Monte Carlo study. We generated a number of datasets to conduct
the evaluation of the clustering methods \textsc{cw} and \textsc{hf} that we
compare. We will examine the performance of these methods to determine
success rates in detecting club membership for various parameter
configurations including the number of countries, club size, time span and
number of clubs. The analysis is carried out for two separate cases. In the
first case we analyze single club data, while in the second case we include
multiple clubs.

Below we present the data generating processes and evaluation procedures
employed in this study. A similar design was used by Beylunio\u{g}lu et al
(2018) in assessing the properties of the maximum clique method, an
alternative clustering mechanism that relies on Augmented Dickey Fuller (%
\textsc{adf}) unit root testing. However, the maximum clique method is a
``top down'' method that leads to a different definition of clubs and is not
considered in the present comparison.\footnote{%
Another method developed by Phillips and Sul (2007) stands out by means of
not requiring a priori classification of countries. However, we exclude this
method for the reason that it is based on the notion of $\sigma $
convergence. The method depends on the definition of convergence by means of
reduction of variance over time and thus convergence of series to a steady
state. Therefore, it is not appropriate to compare this method with HF and
the CW method developed in this study as both of the latter deal with
convergence of the mean (function) of the series.}

\subsubsection*{Data Generating Processes}

The simulation assumes that the log \textsc{gdp} series for region $i$ is
given by
\begin{equation}
y_{it}=\alpha _{i}+d_{i}r_{t}+\epsilon _{it},  \label{dgp}
\end{equation}%
\noindent where $\epsilon _{it}\sim I(0)$ is the error term and $r_{t}$ is
the common factor which affects all countries the same way (such as
technology). If we assume non-stationarity of the factor, a pair of
countries can only be convergent if the country specific constants, $d_{i}$,
which measure the impact of common factor are equal. In other words, for the
pair $i$ and $j$, if ${i}=d_{j}$, $r_{t}$ is canceled out and $y_{it}-y_{jt}$
becomes $\alpha _{i}-\alpha _{j}+\epsilon _{it}-\epsilon _{jt}$. In this
case, since the error terms are assumed to be stationary, we have $\alpha
_{i}-\alpha _{j}+\epsilon _{it}-\epsilon _{jt}\sim I(0)$ and the pair $i$
and $j$ would be convergent by definition. Likewise, for any subset of
countries having equal $d_{i}$, all pairwise difference series in that
subset would be stationary and hence these countries would constitute a
convergence club. Finally, the constants, $\alpha _{i}$ are country specific
and are generated once for all data sets.

The non-stationarity of $r_{t}$ is modeled using the \textsc{arima} process
\begin{equation}
r_{t}=r_{t-1}+v_{t},\quad v_{t}=\rho _{v}v_{t-1}+e_{t},\quad e_{t}\sim iid\
N(0,1-\rho _{v}^{2}),
\end{equation}%
\noindent where we allow $\rho _{v}=\{0.2,0.6\}$ as separate cases. In
addition we also allow the error term of the log \textsc{gdp} series in
equation (\ref{dgp}) to have serial dependence, following the specification
\begin{equation}
\epsilon _{it}=\rho _{i}\epsilon _{i,t-1}+v_{it},\quad v_{it}\sim \
iid\,N(0,\sigma _{v_{i}}^{2}(1-\rho _{i}^{2})),
\end{equation}%
where we assume that the error terms $v_{it}$ are i.i.d. distributed Normal
random variables. The autoregressive coefficient $\rho _{i}$ and $\sigma
_{v_{i}}^{2}$ are country specific and invariant among the single and
multiple clubs datasets. We generated the coefficients to have the following
property.
\begin{equation*}
\sigma _{vi}^{2}\sim \ iid\ \mathcal{U}[0.5,1.5],\quad \rho _{i}\sim \ iid\
\mathcal{U}[0.2,0.6]
\end{equation*}

To generate a dataset containing a single club the coefficients of the $m$
convergent countries are assumed to be $d_{i}=d_{j}=1$. For the remaining ($%
N-m$) countries, $d_{i}$ is generated randomly as $d_{i}\sim iid\ \mathcal{X}%
_{m}^{2}$. Similarly, we also generate country specific constants as $\alpha
_{i}\sim iid\ \mathcal{X}_{m}^{2}$.

For multiple clubs, in order to assess successful detection in club
membership we want to make sure that there are some non-convergent countries
present in the data that do not belong to any club. In that case, the value $%
m$ of club size, when the number of clubs ($k$) and the number of countries (%
$N$) are given, is chosen in such a way as to allow for at least a pair of
non-convergent countries to be present in order to evaluate successful
converging behaviour.
%\todo{\tiny \emph{is determined ... for at least a pair of non-convergent countries to be present}
%Is this clear? Why at least a pair? Answer: The text is rewritten, where "determined" is replaced by "chosen" }
For a given $k$ and $N$, the clubs sizes $m$'s are randomly drawn from a
Poisson distribution with a rate of $\lambda =N/k$. For each $N$, random
draws are repeated $k$ times.\footnote{%
Obviously we did not allow the sum of $m$ to exceed $N$, if this happens we
redraw the last club size.}

The simulations are repeated 2000 times using different combinations of $%
T=\{50,100\}$ time intervals, $N=\{10,20,30\}$ count of countries, $%
m=\{3,5,7,10\}$ number of club members for the single club case. In the
multiple club case we considered $T=\{50,100\},$ $N=10,$ $m=\{3,5\}\ $and $%
k=\{2,3\}$ number of clubs.\footnote{%
The computational burden for larger values of $m$, $k$ and $N$ proved to be
be quite high at this point.}

\subsubsection{Testing and Evaluating Procedures}

To evaluate convergence we utilise evaluation tools from the literature on
forecasting. The first one is the Kupiers Score (\textsc{ks}), while the
second one is the Pesaran and Timmermann (1992) (\textsc{pt}) test statistic
commonly used in the forecasting times series literature for the evaluation
of sign forecasts. It is worth noting that sign forecasts are used for
predicting whether an underlying series would increase relative to a
benchmark such as, for example, a zero return threshold. This test is cast
in terms of a binary process where success is the increase relative to the
chosen benchmark. In our case we take a ``success''\ as the correct
detection of a country's membership in a club. In the context of
forecasting, this is equivalent to success in forecasting the sign of a time
series.

Since granting membership into a club or denying it can occur randomly,%
\footnote{%
This is similar to expecting an unbiased coin to come up heads with 50\%
probability.} \textsc{ks} takes the correct forecasts and false alarms into
account separately. \textsc{ks} is defined as $H-F$ where
%Answer: (The sentence in the comment can be added) :For example, if we are evaluating a forecast of a bad
%event or calamity in economics, an estimate of false alarms would help us
%avoid the issue of scare-mongering.
\begin{equation*}
H=\frac{II}{II+IO},\text{ and }F=\frac{OI}{OI+OO}.
\end{equation*}%
$I$ ($O$) are binary indicators indicating whether the country under
investigation is a member (not a member) of a given club. In considering the
pairs of letters, the first letter indicates whether the country is found to
be a member in the Monte Carlo experiment, while the second letter denotes
its actual membership state (i.e. whether the country is actually in the
club or not). $II$ then indicates that a country as a member of the club is
correctly identified; $OO$ denotes that a country is correctly identified as
not a member of the club. Furthermore, $IO$ indicates that a country is
detected to be a member of the club, while actually it is not (false
detection). $OI$ refers to the reverse case where the country is
misclassified as being outside, even though it is a member of the club
(false alarm). The ratio $H$ captures the rate of ``correct hits'' in
detecting club membership, whereas $F$ denotes the ``false alarm'' rate,
that is the rate of false exclusions.

As in the case of sign prediction in the forecasting literature, success can
be the outcome of a pure chance probability event of 0.5. Hence, to test the
statistical significance of \textsc{ks}, we will employ the following
\textsc{pt} statistic
\begin{equation*}
PT=\frac{\widehat{P}-\widehat{P^{\ast }}}{[\widehat{V}(\widehat{P})-\widehat{
V}(\widehat{P^{\ast }})]^{\frac{1}{2}}}\sim N(0,1).
\end{equation*}
\noindent $\widehat{P}$ refers to the proportion of correct predictions
(correct detections of countries as being a member or non member) over all
predictions ($N$ countries), and $\widehat{P^{\ast }}$ denotes the
proportion of correct detections under the hypothesis that the detections
and actual occurrences are independent (where success is a random event of
probability 0.5). $\widehat{V}(\widehat{P})$ and $\widehat{V}( \widehat{%
P^{\ast }})$ stand for the variances of $\widehat{P}$ and $\widehat{ P^{\ast
}}$ respectively.

In simulations involving multiple clubs, it is not possible to use either
the \textsc{ks} or the \textsc{pt} statistic given that the success/failure
classification is no longer binary - as in the case of the single club case.
In the multiple club case there are more than two distinct cases for the
actual membership state: the country can be either a member of the correct
club, belong to the ``wrong'' club, or not be a member of any club.

To confront this problem, in the case of multiple clubs we utilise a much
stricter criterion by counting the successful cases in our simulations in
which \textit{all} countries are detected correctly. We do not evaluate
success as a binary outcome, country by country as in the case of a single
club in each replication, but we only count as success having all countries
satisfying the convergence condition. This is a much stricter criterion
given that success depends on the overall results in each replication in
which \textit{all} countries are detected correctly.

%Dropped ****
%If, in one replication, \emph{all} countries are placed correctly
%in their correct club we will consider this as one successful outcome out of
%a total of 1000 replications and as a fail otherwise.

\subsection{Simulation Results}

Below we discuss the findings of the simulations based on the data
generating processes of club formation. The comparison involves the
bootstrap version of the \textsc{kpss} test (henceforth \textsc{cw})
proposed in this paper, and the original \textsc{hf} test based on the
asymptotic version of the multivariate \textsc{kpss} test.

\subsubsection{Single Club Results}

The results are presented in Table 1 for 0.05 and 0.10 significance levels.\footnote{%
We also have the results for the 0.01 significance level but to conserve
space we do not report them. They are available from the authors on request.}
The total number of countries $N$ are set at ($N={10,20,30}$); there are two
choices of time span ($T={50,100}$) that mimic the real data time span
availability; and two choices of the persistence parameter ($\rho
_{v}=\{0.2,0.6\}$). It is expected that as the number of countries $N$ and
club size $m$ increase, the likelihood of an incorrect classification will
also increase, but the opposite will be the case for an increase of the time
span $T$ for given $N$ and $m.$

As seen in Table 1, \textsc{cw} outperforms \textsc{hf} in all categories.
For example, with $m=3,$ $N=10,T=50$ and $\rho =0.2$ (configuration $%
\mathcal{A}$), and significance levels 0.05 and 0.10, the \textsc{ks }%
results (the ``correct hit'' ratio net of ``false alarms'') for \textsc{cw} are,
respectively, 0.87 and 0.89. The comparable numbers for \textsc{hf} are 0.63
and 0.66. Similarly, for the cases with $m=10,$ $N=30,T=100$ and $\rho =0.6$
(configuration $\mathcal{B}$), \textsc{cw} with 0.77 and 0.76 outperforms
the \textsc{hf} method - 0.60 and 0.60. The results are in line with our
prior expectations that larger $N$ and $m$ values would result in lower
success rates. However, in all cases the \textsc{cw} test does better.

The \textsc{pt} statistics\footnote{The PT statistic follows an asymptotic standard normal variate.}
for configuration $\mathcal{A}$ yield values 1.91 and 2.02 for \textsc{hf}
and 2.51 and 2.49 for \textsc{cw}; for configuration $\mathcal{B}$ the
\textsc{hf} values are 2.86 and 2.97; with 3.37 and 3.48 for \textsc{cw}.

Note that the rejections of the null hypothesis of random success outcomes
are higher with the \textsc{pt} test for \textsc{cw} in all cases. The
results clearly demonstrate that the \textsc{cw}, bootstrap \textsc{kpss}
test offers a significant improvement over the \textsc{hf} procedure.

\subsubsection{Multiple Club Results}

The results for the multiple club case are presented in Table 2. The
multiple clubs cases involve classifications with $k={2}$ and ${3}$ and $N={%
10}$ for $T=\{{50,100\}}$ and $\rho _{v}=\{0.2,0.6\}$. The club sizes
associated with each club are listed in the second column of Table 2 for
each $k.$ For example, the entry $4,4$ for club size $m$ refers to two clubs
of equal size $4$, for $k=2$. In the case of $k=3,$ $m$ enters as $3,3,2$,
that is two clubs of size $3$ and one club of size $2$.

The \textsc{cw} test outperforms \textsc{hf} in the majority of cases. For example, with $N=10,$ $%
k=2,$ $T=100$ and $\rho _{v}=0.6,$ \textsc{cw} detects 54\%, 45\% and 37\%
correct classifications at the 0.01, 0.05 and 0.10 significance levels;
\textsc{hf} does that with frequency 44\%, 38\% and 34.80\% respectively.
For the case when $k=2$ and  $T=50$, \textsc{hf} does slightly better than
\textsc{cw,} but when $k=3$ the performance of \textsc{hf} deteriorates
rapidly. In that case when $N=10,$ $k=3,$ $T=100$ and $\rho
_{v}=0.2,$ \textsc{cw} detects 27\%, 25\% and 22\% correct classifications,
while the comparable results for \textsc{hf}, respectively, are 10.40\%,
8.20\% and 5.60\%. Since we have adopted a much stricter criterion where
success is defined as \textit{all }countries detected correctly, we do expect lower rates of correct
detection than was the case for single clubs. In all cases, we see an improvement for \textsc{cw} when the number of clubs increases even when $T$ is relatively small, but not for \textsc{hf}.

Overall, the multiple club results suggest that in terms of accuracy the
\textsc{cw} does better in detecting the presence of clubs or clusters of
countries. This gives us confidence that applying the above method to real
data can provide us with useful insights about how countries over time
collect themselves into different club formations of similar characteristics
as far as economic activity is concerned.

\section{Applications}

As shown in the Monte Carlo study a problem with the asymptotic test is that
it does not permit reliable inference with only 30 years of data. Using an
asymptotic test of the null of stationarity (convergence) tends to distort
club membership detection due to size distortions which are ameliorated when
we implement the bootstrap. That is, the test is oversized resulting in a
tendency to reject the null hypothesis of convergence. In this section we
assess the extent to which a size distortion affects our inference on the
degree of convergence using two real-world datasets. We first compare the
results of the asymptotic and bootstrap tests using the cross-country
dataset originally adopted by Hobijn and Franses (2000). We then utilise
data gathered at a finer geographical and sectoral scale by making the same
comparison based upon the European regional dataset used by Corrado, Martin
and Weeks (2005). We expect to resolve the size distortion which afflicts
the asymptotic test and to find more evidence of convergence than what
originally acknowledged by Hobijn and Franses (2000).

\subsection{Cross-Country Convergence}

In this section we compare the asymptotic and the bootstrap results using
the Hobijn and Franses (2000) dataset for the period 1960-1989 which comprises 112 countries from the
Penn World Table listed in Table 3. Focussing upon log per capita \textsc{gdp%
}, the results based upon the asymptotic test have a striking feature,
namely a very large number of convergence clubs. In particular, Hobijn and
Franses (2000) find 63 asymptotically perfect convergence clubs and 42
asymptotically relative\footnote{%
Note that perfect convergence implies convergence to identical log real GDP
per capita levels. Relative convergence implies convergence to constant
relative real GDP per capita levels.} convergence clubs.\footnote{%
These are the results using $p_{\min }=0.01$ and $L=2$ as presented in
Tables BII and BIII of the Hobijn and Franses (2000) paper.} As Table 4
shows, in the case of perfect convergence the lack of convergence is
manifest in 29 singletons and 22 two-country clusters. A similar result can
be observed in the case of relative convergence where Hobijn and Franses
(2000) find a large number of two and three-country clusters. The lack of
convergence is also evident in the fact there are no clusters of size six or
more for asymptotic perfect convergence and only one club of size six for
relative convergence.

We therefore implement the bootstrap version of the test on the same dataset
and find a significant increases in the extent of convergence. Specifically,
in the perfect convergence case and relative to the asymptotic results, we
observe a 57\% reduction in the number of convergence clubs (from
63 to 27); for relative convergence the reduction is 38\% (from 42
to 26). In other words, there is evidence towards finding more convergence.

Looking at the change in the distribution of cluster sizes for perfect
convergence, we observe a dramatic reduction (by 96\%) in the number of
singletons (from 29 to 1) and by 81\% in the number of two-country clusters
(from 22 to 4). Commensurate with this finding, we note that countries are
now clustering at a larger scale with two clubs having up to seven countries
and with a substantial increase in the number of clusters containing five
and six countries. A similar increase in the degree of convergence can be
observed in the case of relative convergence.

Tables 5 and 6 report the asymptotic and bootstrap cluster composition for
relative convergence. A number of noteworthy observations can be made. We
confirm the findings of Hobijn and Franses that convergence is more
widespread among low income economies, and in particular Sub-Saharan Africa.
Similarly we find that in general low income countries do not converge
to high income. The two exceptions to this found in the asymptotic results,
namely Kenya and Ecuador forming clubs alongside Australia and Denmark and
Canada, are not found in the bootstrap results. In contrast to Hobijn and
Franses we do find a significantly higher degree of convergence, both
amongst low income and high income countries. The results based on the
asymptotic test indicate very little convergence for the richer economies
with all groups of size two. The bootstrap test locates a greater degree of
convergence, for example, cluster 11 (Germany, Denmark, France, Luxemburg
and New Zealand) and cluster 17 (Belgium, Great Britain, Netherlands and
Norway). %\fxnote{Highlight USA/Mex?}

In the next section we apply the asymptotic and bootstrap version of the
test to the European regional dataset originally used by Corrado, Martin and
Weeks (2005). A critical difference with respect to the analysis undertaken
at the aggregate country level, is that we allow for the possibility that
convergence is more prevalent at a sector-specific level, and in addition
consider a smaller geographical unit. Much of the theory of convergence
highlights the potential role of technology spillovers as one of the
possible drivers of convergence. As a result, in what follows we move away
from an aggregate analysis to considering how convergence differs across
agricultural, manufacturing and service sectors.

\subsection{European Regional Convergence}

In the following sections we examine the extent of regional convergence
within the \textsc{eu}. Regional convergence -- or what the European
Commission calls `regional cohesion' -- is a primary policy objective, and
is seen as vital to the success of key policy objectives, such as the single
market, monetary union, \textsc{eu} competitiveness, and enlargement
(European Commission, 2004). As a result, the theory of and evidence on
long-run trends in regional per capita incomes and output are of critical
relevance to the \textsc{eu} regional convergence and regional policy debate
(Boldrin and Canova, 2001). Indeed, according to Fujita \textit{et al.}
(1999), the implications of increasing economic integration for the \textsc{%
eu} regions has been one of the factors behind the development of the `new
economic geography' models of regional growth. To date, however, very few of
these models have been tested empirically on \textsc{eu} evidence.

In response to the policy and research questions outlined above our
empirical analysis will be framed around the identification of regional
convergence clubs in the \textsc{eu}. To identify regional convergence
clusters we use the method introduced by Hobijn and Franses (2000) which
allows for the endogenous identification of the number and membership of
regional convergence clusters (or `clubs') and compare the results of the
bootstrap and asymptotic versions of the test to assess the differences in
terms of number, size and composition of the resultant clusters.

\subsubsection{Data}

The so-called Nomenclature of Statistical Territorial Units (\textsc{nuts})
subdivides the economic territory of the 15 countries of the European Union
using three regional and two local levels. The three regional levels are:
\textsc{nuts}3, consisting of 1031 regions; \textsc{nuts}2, consisting of
206 regions; and \textsc{nuts}1 consisting of 77 regions. \textsc{nuts}0
represents the delineation at the national level and comprises France,
Italy, Spain, \textsc{uk}, Ireland, Austria, Netherlands, Belgium,
Luxemburg, Sweden, Norway, Portugal, Greece, Finland, Denmark and West
Germany. We are aware of the problems that surround the choice of which
spatial units to use.\footnote{%
Chesire and Magrini (2000) provide a useful discussion of these issues,
focussing on the importance of centering the analysis on regions that are
self-contained in labour market terms.} For example, many of the regional
units used by \textsc{eurostat} have net inflows of commuters and in
addition, these regions also tend to be those with the highest per capita
income. Boldrin and Canova (2001) criticize the European Commission for
utilizing inappropriate regional units. Whereas \textsc{nuts}1, \textsc{nuts}%
2 and \textsc{nuts}3 regions are neither uniformly large or sufficiently
heterogeneous such that a finding of income divergence across regions cannot
unequivocally be taken as evidence for the existence of an endogenous
cumulative growth processes. In fact, the smaller the geographical scale,
the more incomplete and fragmented is the statistical information we can
get. Although we do not wish to detract from the importance of these
matters, in this study our primary focus is a comparison of two different
tests for regional convergence for which the unit of analysis is the same.
In conducting our analysis we choose to focus on \textsc{nuts}1 regions,
achieving a compromise between the availability of reliable data at a
regional level which is sufficiently homogeneous, and the need to move
beyond national borders. The complete list of \textsc{nuts}1 regions%
\footnote{%
For Portugal, Luxemburg and Ireland, data are only available at the \textsc{%
nuts}0 level. For Norway we have no data at the \textsc{nuts}1 level. Time
series data for the sample period considered are not available for East
Germany, which is therefore excluded from the analysis.} used in this study
is given in Table 7.

We use regional data on Gross Value Added\footnote{\textsc{gva} has the
comparative advantage with respect to \textsc{gdp} per capita of being the
direct outcome of various factors that determine regional competitiveness.
Regional data on \textsc{gva} per-capita at the \textsc{nuts}1 level for
agriculture, manufacturing, market and non-market services, have been kindly
supplied by Cambridge Econometrics, and are taken from their European
Regional Database. All series have been converted to constant 1985 prices (%
\textsc{ecu}) using the purchasing power parity exchange rate.} per worker
for the period 1975 to 1999 for the agriculture, manufacturing and services
sectors. Although data are available for more recent years, we focus on this
particular time frame to facilitate a comparison with the results of
Corrado, Martin and Weeks (2005). The service sector has been further
sub-divided into market and non-market services: market services comprise
distribution, retail, banking, and consultancy; non-market services comprise
education, health and social work, defence and other government services.

\subsubsection{Results}

In this section we present the main results of our analysis. Given the large
number of \textsc{eu} regions in Figures 1 and 2 we first present the
results for the asymptotic and bootstrap test of convergence in mapped
rather than tabular form. Table 8 summarises this information in terms of
the number and size of the convergence clubs and group characteristics, such
as average per-capita income.

\subsubsection{Graphing Convergence Clusters}

In Figures 1 and 2 clusters which contain the largest number of member
regions are indicated with a darker shade on each map. Regions which belong
to two-region clusters or do not cluster with any other region have no
shading. In the key to the maps, the first number indicates the cluster size
and the second letter denotes the cluster identifier. In Figure 1 maps a)
and b) ( c) and d)) present the asymptotic and bootstrap generated outcomes
for agriculture (manufacturing). The relative pattern of convergence
corroborates with our prior expectations, namely that the bootstrap test is
obviously rejecting the stationary null with a lower frequency and thereby
locating more evidence for convergence. In Figure 2 we find a similar
pattern for market and non-market services.

In Table 8 we present the frequency distribution of the cluster size for
both bootstrap and asymptotic tests and for each\footnote{%
In order to directly compare the bootstrap and asymptotic results in Corrado
et al. (2005) we set $p_{\min }$ to be equal to 0.01 and the bandwidth $L=2$%
. The number of bootstrap samples is set at 200.} economic sector. Row
totals provide an indication of the degree of convergence for each economic
sector. Column totals provide information on the number of convergence clubs
across sectors by cluster size. The asymptotic results are displayed in
panel I and the bootstrap results are displayed in panel II. Overall, we
observe a common pattern, namely a shift in the probability distribution
towards a fewer number of clusters of larger size, and a commensurate
increase in the extent of regional convergence. The total number of clusters
for the asymptotic tests is 81, which falls by 32\% to 55 clusters for the
bootstrap test. This pattern is repeated for all sectors. Comparing column
totals across the two tests is also informative since it gives the total
number of clusters by cluster size, also shown in Figure 3. For the
asymptotic test, more than 80\% of the probability mass is distributed in
clusters of size 4 or less, with approximately 10\% of clusters of size 6 or
more. In contrast, for the bootstrap test, approximately 50\% of the
clusters have a cluster size of 4 or less, with approximately 40\% of
clusters of size 6 or more.

Examining the results for each sector, for agriculture the size of the
largest cluster generated by bootstrap critical values increases from seven
to ten regions, with a commensurate decrease in the number of clusters of
size 5 or less. Similarly for the manufacturing sector we observe an
increase in the size of the largest cluster from six to nine regions and a
decrease in the number of clusters of size 4 or less. In the market-service
sector there is a reduction in the size of the largest cluster from nine to
eight and for non-market services there is no change in the size of the
largest cluster, but a substantial increase in clustering at the medium
scale. In both service sectors there is a decrease in the number of clusters
of size 4 or less.

\textbf{Cluster Composition }In establishing whether the composition of the
clusters (i.e. the constituent regions) is changing between the two tests,
we first collect the asymptotic (A) generated cluster outcomes in a $N\times
N$ matrix $\mathbf{M}^{A}=\{m_{ij}^{A}\};$ element $m_{ij}^{A}$ equals to 1
if regions $i$ and $j$ belong to the same cluster and zero otherwise. $%
\mathbf{M}^{B}=\{m_{ij}^{B}\}$ denotes the same for the bootstrap (B)
generated cluster outcomes. The correlation parameter between the
asymptotic, $\mathbf{M}^{A}$, and the bootstrap cluster pattern, $\mathbf{M}%
^{B},$ is then given by
\begin{equation}
\zeta _{l}=\left( \frac{\sum\limits_{i=1}^{N}\sum\limits_{j\neq
i}^{N}m_{ij}^{B}\times m_{ij}^{A}}{\left(
\sum\limits_{i=1}^{N}\sum\limits_{j\neq i}^{N}m_{ij}^{B}\right) ^{1/2}\left(
\sum\limits_{i=1}^{N}\sum\limits_{j\neq i}^{N}m_{ij}^{A}\right) ^{1/2}}%
\right) ^{1/2},
\end{equation}%
where $l$ indexes the set \{Agriculture, Manufacturing, Market
Services, Non-Market Services\}. The results are reported in panel III of
Table 8. With correlation coefficients ranging between 50\% for
manufacturing and 67\% \ for agriculture, we note further evidence of a
significant difference in the composition of the clusters generated by the
asymptotic and bootstrap tests.\footnote{%
The method used in this paper to locate convergence clubs bypasses the
particular problem of exactly how to utilize conditioning information in the
model specification. Corrado and Weeks (2011) provide further information on
how to interpret the results by confronting the resulting cluster
composition, for both the asymptotic and the bootstrap tests, with a set of
hypothetical clusters based on different theories and models of regional
growth and development.}

\textbf{Mean Income }In order to assess the properties of each cluster we
compute mean log per-capita income,\footnote{%
Mean income is the cluster mean of log per-capita GVA.} $\bar{y}_{g}$ for
each test. The top panel of Figure 4 shows that the asymptotic test
generates a distribution with a large number of small clubs while in the
bootstrap test there are a fewer number of clusters of larger size. A visual
impression of the oversized property of the asymptotic test of convergence
is also evident in the distribution of the cluster mean of log per-capita
income and in a relatively higher right kurtosis of this distribution, as
presented in the lower panel of Figure 4. In this case an overrejection of
the convergent null generates a distribution with a large number of small
clubs characterised by a higher mean log per-capita income which results in
a widening of the gap between the poorest and the richest clusters. In
examining the comparable bootstrap distribution we observe a marked decrease
in right kurtosis and a commensurate narrowing of the gap between the
richest and the poorest cluster. Summary statistics are provided in the last
three columns of panels I and II of Table 8. Note that for the bootstrap
distribution the reduction in the gap between the richest and the poorest
clusters is evident in a lower standard deviation of mean cluster per-capita
income (from 15.2 to 5.4). The narrowing of the gap between the richest and
poorest cluster translates into an increase in mean log per-capita income of
the \emph{poorest} cluster, $\bar{y}_{min}$, by around 24\% (from 9.4 to
11.7) and a decrease in mean log per-capita income of the \emph{richest}
cluster, $\bar{y}_{max},$ by almost 50\% (from 103 to 62.6). These results
demonstrate the importance of the correct identification of convergence
clubs. Given that many policy instruments are designed to reduce the gap
between the richest and the poorest regions, basing inference and policy
decisions on the results of the asymptotic test would indicate the need for
a stronger action than is actually needed when looking at the bootstrap test
outcomes.

\section{Conclusions}

This study represents an extension of the multivariate test of stationarity
which allows for endogenous identification of the number and composition of
regional convergence clusters using sequential pairwise tests for
stationarity. The main drawback of this approach is the short time-horizon
which affects the size of the test. Our proposed bootstrap based extension
to the sequential pairwise multivariate tests for stationarity performs well
in Monte Carlo simulations in identifying and detecting correctly cluster
membership when compared with the asymptotic version of the Hobijn and
Franses (2000) approach. Based upon Monte Carlo evidence comparing the
performance of \textsc{cw} with \textsc{hf} varying the number of countries,
data span, club size and degree of persistence, indicate that detection
rates of club membership (net of misclassifications) improve considerably
when we implement the bootstrap. In operationalizing a bootstrap test of
multivariate stationarity our results confirm the oversized property of the
asymptotic test, and reveal a significantly greater degree of convergence.
This evidence is gathered using both cross-country and regional data for the
European Union for a number of industrial sectors. Our results show that by
resolving the size distortion which afflicts the asymptotic test we find
considerably more evidence of convergence in both the applications
considered.

\newpage

\pagestyle{plain}

\pagebreak

\begin{table}[ht]
\caption{Single Clubs Results}%\centering
\begin{adjustwidth}{-1.cm}{}
		\centering
		\begin{tabular}{|l|lll|lrr|lrr|lrr|lrr|}
			\hline
			\multicolumn{4}{|c|}{DataType} &  \multicolumn{6}{|c|}{KS} &  \multicolumn{6}{|c|}{PT}   \\
			\hline
			N & m & $\rho$ & T & \multicolumn{3}{|c|}{HF}  &  \multicolumn{3}{|c|}{CW}  &  \multicolumn{3}{|c|}{HF}  &   \multicolumn{3}{|c|}{CW} \\
			&  &  &  & 0.01 & 0.05 & 0.10 & 0.01 & 0.05 & 0.10 & 0.01 & 0.05 & 0.10 & 0.01 & 0.05 & 0.10  \\  \cline{1-16}
			\multirow{8}{*}{10} & \multirow{4}{*}{3} & 0.2 & 50 &  0.60 & 0.63 & 0.66 & 0.85 & 0.87 & 0.89 & 1.67 & 1.91 & 2.02 & 2.49 & 2.51 & 2.72 \\
			&  &  & 100 & 0.73 & 0.78 & 0.83 & 0.76 & 0.79 & 0.81 & 1.96 & 2.16 & 2.34 & 2.26 & 2.28 & 2.49 \\
			&  & 0.6 & 50 & 0.69 & 0.71 & 0.74 & 0.87 & 0.88 & 0.91 & 2.04 & 2.17 & 2.40 & 2.51 & 2.58 & 2.64 \\
			&  &  & 100 & 0.73 & 0.81 & 0.82 & 0.77 & 0.81 & 0.81 & 2.02 & 2.42 & 2.52 & 2.29 & 2.33 & 2.29 \\  \cline{2-4}
			& \multirow{4}{*}{5} & 0.2 & 50 & 0.61 & 0.61 & 0.64 & 0.87 & 0.88 & 0.87 & 1.68 & 2.00 & 1.95 & 2.68 & 2.73 & 2.69 \\
			&  &  & 100 &0.69 & 0.76 & 0.73 & 0.79 & 0.82 & 0.82 & 2.10 & 2.18 & 2.31 & 2.62 & 2.68 & 2.69 \\
			&  & 0.6 & 50 &  0.67 & 0.67 & 0.67 & 0.89 & 0.89 & 0.88 & 2.15 & 2.25 & 2.21 & 2.75 & 2.76 & 2.73 \\
			&  &  & 100 &0.71 & 0.76 & 0.77 & 0.80 & 0.83 & 0.84 & 2.21 & 2.40 & 2.40 & 2.62 & 2.63 & 2.69 \\  \cline{1-4}
			\multirow{16}{*}{20} & \multirow{4}{*}{3} & 0.2 & 50 & 0.47 & 0.47 & 0.57 & 0.66 & 0.74 & 0.73 & 1.60 & 2.18 & 2.43 & 2.87 & 3.08 & 3.53 \\
			&  &  & 100 & 0.60 & 0.58 & 0.63 & 0.54 & 0.61 & 0.65 & 2.39 & 2.46 & 2.71 & 2.09 & 2.39 & 2.98 \\
			&  & 0.6 & 50 & 0.58 & 0.60 & 0.64 & 0.72 & 0.81 & 0.81 & 1.82 & 1.98 & 2.35 & 3.14 & 3.46 & 3.51 \\
			&  &  & 100 &  0.65 & 0.69 & 0.73 & 0.62 & 0.72 & 0.72 & 2.64 & 2.83 & 2.91 & 2.13 & 2.57 & 3.21 \\  \cline{2-4}
			& \multirow{4}{*}{5} & 0.2 & 50 & 0.60 & 0.58 & 0.58 & 0.87 & 0.88 & 0.87 & 2.74 & 2.88 & 2.94 & 3.84 & 3.86 & 3.82 \\
			&  &  & 100 & 0.67 & 0.71 & 0.71 & 0.80 & 0.82 & 0.83 & 2.92 & 2.99 & 3.04 & 3.40 & 3.56 & 3.75 \\
			&  & 0.6 & 50 & 0.63 & 0.63 & 0.66 & 0.92 & 0.91 & 0.90 & 2.34 & 2.56 & 2.55 & 3.91 & 3.89 & 3.72 \\
			&  &  & 100 &0.71 & 0.70 & 0.71 & 0.81 & 0.82 & 0.84 & 3.05 & 3.39 & 3.17 & 3.48 & 3.37 & 3.62 \\  \cline{2-4}
			& \multirow{4}{*}{7} & 0.2 & 50 &0.61 & 0.60 & 0.58 & 0.90 & 0.87 & 0.86 & 2.78 & 2.80 & 2.74 & 3.98 & 3.79 & 3.79 \\
			&  &  & 100 & 0.64 & 0.66 & 0.66 & 0.88 & 0.88 & 0.88 & 3.03 & 2.90 & 3.09 & 4.03 & 4.01 & 4.11 \\
			&  & 0.6 & 50 &0.63 & 0.61 & 0.62 & 0.91 & 0.88 & 0.86 & 2.13 & 2.67 & 2.68 & 4.06 & 3.90 & 3.83 \\
			&  &  & 100 &  0.68 & 0.69 & 0.68 & 0.88 & 0.88 & 0.89 & 3.22 & 3.08 & 3.29 & 3.91 & 3.85 & 3.94 \\   \cline{2-4}
			& \multirow{4}{*}{10} & 0.2 & 50 & 0.52 & 0.48 & 0.47 & 0.73 & 0.72 & 0.70 & 2.42 & 2.48 & 2.37 & 3.28 & 3.25 & 3.24 \\
			&  &  & 100 & 0.61 & 0.64 & 0.64 & 0.81 & 0.79 & 0.79 & 3.09 & 3.06 & 3.01 & 3.83 & 3.70 & 3.64 \\
			&  & 0.6 & 50 &0.56 & 0.50 & 0.52 & 0.76 & 0.74 & 0.74 & 2.24 & 2.18 & 2.28 & 3.36 & 3.33 & 3.36 \\
			&  &  & 100 & 0.65 & 0.62 & 0.62 & 0.82 & 0.80 & 0.79 & 3.08 & 3.09 & 3.02 & 3.77 & 3.62 & 3.61 \\  \cline{1-4}
			\multirow{16}{*}{30} & \multirow{4}{*}{3} & 0.2 & 50 & 0.54 & 0.49 & 0.56 & 0.76 & 0.80 & 0.78 & 1.76 & 1.85 & 1.97 & 2.56 & 2.65 & 2.71 \\
			&  &  & 100 & 0.59 & 0.62 & 0.65 & 0.72 & 0.76 & 0.77 & 2.11 & 2.33 & 2.56 & 2.21 & 2.31 & 2.38 \\
			&  & 0.6 & 50 & 0.60 & 0.56 & 0.61 & 0.80 & 0.84 & 0.81 & 2.01 & 2.13 & 2.23 & 2.62 & 2.69 & 2.77 \\
			&  &  & 100 & 0.65 & 0.67 & 0.72 & 0.73 & 0.76 & 0.77 & 2.13 & 2.43 & 2.55 & 2.25 & 2.38 & 2.37 \\   \cline{2-4}
			& \multirow{4}{*}{5} & 0.2 & 50 & 0.54 & 0.52 & 0.50 & 0.81 & 0.82 & 0.78 & 1.93 & 1.93 & 2.03 & 2.77 & 2.77 & 2.75 \\
			&  &  & 100 & 0.59 & 0.61 & 0.63 & 0.76 & 0.77 & 0.76 & 2.22 & 2.40 & 2.31 & 2.55 & 2.62 & 2.62 \\
			&  & 0.6 & 50 & 0.56 & 0.54 & 0.57 & 0.85 & 0.85 & 0.85 & 2.13 & 2.13 & 2.14 & 2.83 & 2.83 & 2.79 \\
			&  &  & 100 & 0.64 & 0.67 & 0.66 & 0.79 & 0.79 & 0.80 & 2.29 & 2.40 & 2.43 & 2.57 & 2.66 & 2.67 \\   \cline{2-4}
			& \multirow{4}{*}{7} & 0.2 & 50 & 0.48 & 0.46 & 0.44 & 0.77 & 0.76 & 0.73 & 1.67 & 1.71 & 2.16 & 2.62 & 2.98 & 2.97 \\
			&  &  & 100 & 0.60 & 0.57 & 0.57 & 0.82 & 0.82 & 0.81 & 2.06 & 2.09 & 2.29 & 1.94 & 2.20 & 2.35 \\
			&  & 0.6 & 50 &0.52 & 0.53 & 0.54 & 0.84 & 0.82 & 0.80 & 2.09 & 2.21 & 2.41 & 2.94 & 3.32 & 3.32 \\
			&  &  & 100 &0.63 & 0.64 & 0.62 & 0.85 & 0.85 & 0.85 & 2.22 & 2.49 & 2.75 & 2.24 & 2.62 & 2.65 \\   \cline{2-4}
			& \multirow{4}{*}{10} & 0.2 & 50 &0.43 & 0.41 & 0.41 & 0.67 & 0.67 & 0.65 & 2.46 & 2.44 & 2.49 & 3.76 & 3.82 & 3.80 \\
			&  &  & 100 & 0.53 & 0.52 & 0.53 & 0.74 & 0.73 & 0.73 & 2.68 & 2.94 & 3.00 & 3.24 & 3.34 & 3.40 \\
			&  & 0.6 & 50 & 0.49 & 0.42 & 0.43 & 0.72 & 0.71 & 0.69 & 2.63 & 2.67 & 2.89 & 3.98 & 3.98 & 3.94 \\
			&  &  & 100 & 0.61 & 0.60 & 0.60 & 0.78 & 0.77 & 0.76 & 2.85 & 2.86 & 2.97 & 3.31 & 3.37 & 3.48 \\
			\hline
		\end{tabular}
			
	\end{adjustwidth}
\end{table}

\FloatBarrier

\pagebreak

\begin{table}[ht]
\caption{Multiple Club Results}\centering
\begin{tabular}{|l|llll|lll|lll|}
\hline
\multicolumn{5}{|c|}{Data Type} &  &  &  &  &  &  \\ \hline
N & m & k & $\rho$ & T & \multicolumn{3}{|c|}{HF} & \multicolumn{3}{|c|}{CW}
\\
&  &  &  &  & 0.01 & 0.05 & 0.1 & 0.01 & 0.05 & 0.1 \\ \cline{1-11}
\multirow{8}{*}{10} & \multirow{4}{*}{4,4} & \multirow{4}{*}{2} & 0.2 & 50 &
34.20\% & 28.00\% & 17.00\% & 28.00\% & 22.00\% & 14.40\% \\
&  &  &  & 100 & 51.80\% & 39.00\% & 33.00\% & 61.20\% & 45.00\% & 40.00\%
\\
&  &  & 0.6 & 50 & 38.80\% & 26.60\% & 19.00\% & 35.00\% & 24.00\% & 17.00\%
\\
&  &  &  & 100 & 44.00\% & 38.00\% & 34.80\% & 54.00\% & 45.00\% & 37.00\%
\\ \cline{2-11}
& \multirow{4}{*}{3,3,2} & \multirow{4}{*}{3} & 0.2 & 50 & 1.20\% & 1.00\% &
1.00\% & 15.60\% & 15.60\% & 14.40\% \\
&  &  &  & 100 & 10.40\% & 8.20\% & 5.60\% & 27.00\% & 25.00\% & 22.00\% \\
&  &  & 0.6 & 50 & 13.00\% & 14.00\% & 10.00\% & 13.00\% & 15.60\% & 14.40\%
\\
&  &  &  & 100 & 3.20\% & 2.60\% & 1.00\% & 20.00\% & 17.80\% & 15.00\% \\
\hline
\end{tabular}%
\end{table}

\begin{center}
\begin{table}[tbp]
\caption{List of Countries (PWT)}\centering
\par
{\normalsize
\begin{tabular}{llllll}
\hline\hline
& {\footnotesize Country} &  & {\footnotesize Country} &  & {\footnotesize %
Country} \\ \hline
{\footnotesize AGO} & {\footnotesize Angola} & {\footnotesize GIN} &
{\footnotesize Guinea} & {\footnotesize NLD} & {\footnotesize Netherlands}
\\
{\footnotesize ARG} & {\footnotesize Argentina} & {\footnotesize GMB} &
{\footnotesize Gambia} & {\footnotesize NOR} & {\footnotesize Norway} \\
{\footnotesize AUS} & {\footnotesize Australia} & {\footnotesize GNB} &
{\footnotesize Guinea Bissau} & {\footnotesize NZL} & {\footnotesize New
Zealand} \\
{\footnotesize AUT} & {\footnotesize Austria} & {\footnotesize GRC} &
{\footnotesize Greece} & {\footnotesize PAK} & {\footnotesize Pakistan} \\
{\footnotesize BDI} & {\footnotesize Burundi} & {\footnotesize GTM} &
{\footnotesize Guatemala} & {\footnotesize PAN} & {\footnotesize Panama} \\
{\footnotesize BEL} & {\footnotesize Belgium} & {\footnotesize GUY} &
{\footnotesize Guyana} & {\footnotesize PER} & {\footnotesize Peru} \\
{\footnotesize BEN} & {\footnotesize Benin} & {\footnotesize HKG} &
{\footnotesize Hong Kong} & {\footnotesize PHL} & {\footnotesize Phillipines}
\\
{\footnotesize BGD} & {\footnotesize Bangladesh} & {\footnotesize HND} &
{\footnotesize Honduras} & {\footnotesize PNG} & {\footnotesize Papua N.
Guinea} \\
{\footnotesize BOL} & {\footnotesize Bolivia} & {\footnotesize HTI} &
{\footnotesize Haiti} & {\footnotesize PRI} & {\footnotesize Puerto Rico} \\
{\footnotesize BRA} & {\footnotesize Brazil} & {\footnotesize HVO} &
{\footnotesize Burkina Faso} & {\footnotesize PRT} & {\footnotesize Portugal}
\\
{\footnotesize BRB} & {\footnotesize Barbados} & {\footnotesize IDN} &
{\footnotesize Indonesia} & {\footnotesize PRY} & {\footnotesize Paraguay}
\\
{\footnotesize BUR} & {\footnotesize Myanmar} & {\footnotesize IND} &
{\footnotesize India} & {\footnotesize RWA} & {\footnotesize Rwanda} \\
{\footnotesize BWA} & {\footnotesize Botswana} & {\footnotesize IRL} &
{\footnotesize Ireland} & {\footnotesize SEN} & {\footnotesize Senegal} \\
{\footnotesize CAF} & {\footnotesize Central African Rep.} & {\footnotesize %
IRN} & {\footnotesize Iran} & {\footnotesize SGP} & {\footnotesize Singapore}
\\
{\footnotesize CAN} & {\footnotesize Canada} & {\footnotesize ISL} &
{\footnotesize Iceland} & {\footnotesize SLV} & {\footnotesize El Salvador}
\\
{\footnotesize CHE} & {\footnotesize Switzerland} & {\footnotesize ISR} &
{\footnotesize Israel} & {\footnotesize SOM} & {\footnotesize Somalia} \\
{\footnotesize CHL} & {\footnotesize Chile} & {\footnotesize ITA} &
{\footnotesize Italy} & {\footnotesize SUR} & {\footnotesize Suriname} \\
{\footnotesize CIV} & {\footnotesize Ivory Coast} & {\footnotesize JAM} &
{\footnotesize Jamaica} & {\footnotesize SWE} & {\footnotesize Sweden} \\
{\footnotesize CMR} & {\footnotesize Cameroon} & {\footnotesize JOR} &
{\footnotesize Jordan} & {\footnotesize SWZ} & {\footnotesize Swaziland} \\
{\footnotesize COG} & {\footnotesize Congo} & {\footnotesize JPN} &
{\footnotesize Japan} & {\footnotesize SYC} & {\footnotesize Seychelles} \\
{\footnotesize COL} & {\footnotesize Colombia} & {\footnotesize KEN} &
{\footnotesize Kenya} & {\footnotesize SYR} & {\footnotesize Syria} \\
{\footnotesize CPV} & {\footnotesize Cape Verde Is.} & {\footnotesize KOR} &
{\footnotesize Korea} & {\footnotesize TCD} & {\footnotesize Tcad} \\
{\footnotesize CRI} & {\footnotesize Costa Rica} & {\footnotesize LKA} &
{\footnotesize Sri Lanka} & {\footnotesize TGO} & {\footnotesize Togo} \\
{\footnotesize CSK} & {\footnotesize Czechoslovakia} & {\footnotesize LSO} &
{\footnotesize Lesotho} & {\footnotesize THA} & {\footnotesize Thailand} \\
{\footnotesize CYP} & {\footnotesize Cyprus} & {\footnotesize LUX} &
{\footnotesize Luxembourg} & {\footnotesize TTO} & {\footnotesize %
Trinidad/Tobago} \\
{\footnotesize DEU} & {\footnotesize West Germany} & {\footnotesize MAR} &
{\footnotesize Morocco} & {\footnotesize TUN} & {\footnotesize Tunisia} \\
{\footnotesize DNK} & {\footnotesize Denmark} & {\footnotesize MDG} &
{\footnotesize Madagascar} & {\footnotesize TUR} & {\footnotesize Turkey} \\
{\footnotesize DOM} & {\footnotesize Dominican Rep.} & {\footnotesize MEX} &
{\footnotesize Mexico} & {\footnotesize UGA} & {\footnotesize Uganda} \\
{\footnotesize DZA} & {\footnotesize Algeria} & {\footnotesize MLI} &
{\footnotesize Mali} & {\footnotesize URY} & {\footnotesize Uruguay} \\
{\footnotesize ECU} & {\footnotesize Ecuador} & {\footnotesize MLT} &
{\footnotesize Malta} & {\footnotesize USA} & {\footnotesize United States}
\\
{\footnotesize EGY} & {\footnotesize Egypt} & {\footnotesize MOZ} &
{\footnotesize Mozambique} & {\footnotesize VEN} & {\footnotesize Venezuela}
\\
{\footnotesize ESP} & {\footnotesize Spain} & {\footnotesize MRT} &
{\footnotesize Mauritania} & {\footnotesize YUG} & {\footnotesize Former
Yugoslavia} \\
{\footnotesize FIN} & {\footnotesize Finland} & {\footnotesize MUS} &
{\footnotesize Mauritius} & {\footnotesize ZAF} & {\footnotesize South Africa%
} \\
{\footnotesize FJI} & {\footnotesize Fiji} & {\footnotesize MWI} &
{\footnotesize Malawi} & {\footnotesize ZAR} & {\footnotesize Zaire} \\
{\footnotesize FRA} & {\footnotesize France} & {\footnotesize MYS} &
{\footnotesize Malaysia} & {\footnotesize ZMB} & {\footnotesize Zambia} \\
{\footnotesize GAB} & {\footnotesize Gabon} & {\footnotesize NAM} &
{\footnotesize Namibia} & {\footnotesize ZWE} & {\footnotesize Zimbabwe} \\
{\footnotesize GBR} & {\footnotesize United Kingdom} & {\footnotesize NER} &
{\footnotesize Niger} &  &  \\
{\footnotesize GHA} & {\footnotesize Ghana} & {\footnotesize NGA} &
{\footnotesize Nigeria} &  &  \\ \hline\hline
\end{tabular}%
}
\end{table}
\end{center}

\pagebreak

\begin{center}
\begin{table}[tbp]
\caption{Joint Frequency Distribution (PWT)}\centering
\par
{\footnotesize
\begin{tabular}{llllllllll}
\multicolumn{10}{l}{} \\ \hline\hline
&  &  &  &  &  &  &  &  &  \\
& \multicolumn{7}{c}{I: Asymptotic} &  &  \\
&  & \multicolumn{6}{l}{Number of Clusters} &  &  \\
& \multicolumn{7}{c}{} &  &  \\
Cluster size & \multicolumn{1}{r}{1} & \multicolumn{1}{r}{2} &
\multicolumn{1}{r}{3} & \multicolumn{1}{r}{4} & \multicolumn{1}{r}{5} &
\multicolumn{1}{r}{6} & \multicolumn{1}{r}{7} &  &  \\ \hline
&  &  &  &  &  &  &  &  & Total Clusters \\
Perfect & \multicolumn{1}{r}{29} & \multicolumn{1}{r}{22} &
\multicolumn{1}{r}{9} & \multicolumn{1}{r}{3} & \multicolumn{1}{r}{0} &
\multicolumn{1}{r}{0} & \multicolumn{1}{r}{0} &  & 63 \\
Relative & 2 & 21 & 12 & 4 & 2 & 1 & 0 &  & 42 \\
& \multicolumn{1}{r}{} & \multicolumn{1}{r}{} & \multicolumn{1}{r}{} &
\multicolumn{1}{r}{} & \multicolumn{1}{r}{} & \multicolumn{1}{r}{} &
\multicolumn{1}{r}{} &  &  \\
& \multicolumn{7}{c}{II: Bootstrap} &  &  \\
&  & \multicolumn{6}{l}{Number of Clusters} &  &  \\
&  &  &  &  &  &  &  &  &  \\
Cluster size & \multicolumn{1}{r}{1} & \multicolumn{1}{r}{2} &
\multicolumn{1}{r}{3} & \multicolumn{1}{r}{4} & \multicolumn{1}{r}{5} &
\multicolumn{1}{r}{6} & \multicolumn{1}{r}{7} &  &  \\ \hline
&  &  &  &  &  &  &  &  & Total Clusters \\
Perfect & \multicolumn{1}{r}{1} & \multicolumn{1}{r}{4} & \multicolumn{1}{r}{
4} & \multicolumn{1}{r}{8} & \multicolumn{1}{r}{3} & \multicolumn{1}{r}{5} &
\multicolumn{1}{r}{2} &  & 27 \\
Relative & \multicolumn{1}{r}{1} & \multicolumn{1}{r}{3} &
\multicolumn{1}{r}{4} & \multicolumn{1}{r}{7} & \multicolumn{1}{r}{3} &
\multicolumn{1}{r}{6} & \multicolumn{1}{r}{2} &  & 26 \\
&  &  &  &  &  &  &  &  &  \\ \hline\hline
\end{tabular}%
}
\end{table}

\begin{table}[tbp]
\caption{Asymptotic: Relative Convergence (PWT)}\centering
\par
\begin{tabular}{rllrrrr}
\hline\hline
{\footnotesize No} & \multicolumn{2}{l}{\footnotesize Countries} &  &  &  &
\\ \hline
{\footnotesize 1} & {\footnotesize AUS} & {\footnotesize DNK} &
\multicolumn{1}{l}{\footnotesize KEN} & \multicolumn{1}{l}{\footnotesize LUX}
& {\footnotesize MUS} & {\footnotesize ZAF} \\
{\footnotesize 2} & {\footnotesize AUT} & {\footnotesize ESP} &
\multicolumn{1}{l}{\footnotesize ISR} & \multicolumn{1}{l}{\footnotesize ITA}
& {\footnotesize PRI} &  \\
{\footnotesize 3} & {\footnotesize CAN} & {\footnotesize CSK} &
\multicolumn{1}{l}{\footnotesize ECU} & \multicolumn{1}{l}{\footnotesize GRC}
& {\footnotesize IRL} &  \\
{\footnotesize 4} & {\footnotesize BDI} & {\footnotesize HVO} &
\multicolumn{1}{l}{\footnotesize MLI} & \multicolumn{1}{l}{\footnotesize MWI}
&  &  \\
{\footnotesize 5} & {\footnotesize BGD} & {\footnotesize BUR} &
\multicolumn{1}{l}{\footnotesize HND} & \multicolumn{1}{l}{\footnotesize NZL}
&  &  \\
{\footnotesize 6} & {\footnotesize CAF} & {\footnotesize IND} &
\multicolumn{1}{l}{\footnotesize NER} & \multicolumn{1}{l}{\footnotesize UGA}
&  &  \\
{\footnotesize 7} & {\footnotesize GUY} & {\footnotesize JOR} &
\multicolumn{1}{l}{\footnotesize SLV} & \multicolumn{1}{l}{\footnotesize SYC}
&  &  \\
{\footnotesize 8} & {\footnotesize AGO} & {\footnotesize GHA} &
\multicolumn{1}{l}{\footnotesize HTI} & \multicolumn{1}{l}{} &  &  \\
{\footnotesize 9} & {\footnotesize BEN} & {\footnotesize GIN} &
\multicolumn{1}{l}{\footnotesize VEN} & \multicolumn{1}{l}{} &  &  \\
{\footnotesize 10} & {\footnotesize BOL} & {\footnotesize LKA} &
\multicolumn{1}{l}{\footnotesize PNG} & \multicolumn{1}{l}{} &  &  \\
{\footnotesize 11} & {\footnotesize BRB} & {\footnotesize IDN} &
\multicolumn{1}{l}{\footnotesize THA} & \multicolumn{1}{l}{} &  &  \\
{\footnotesize 12} & {\footnotesize CIV} & {\footnotesize COG} &
\multicolumn{1}{l}{\footnotesize MAR} & \multicolumn{1}{l}{} &  &  \\
{\footnotesize 13} & {\footnotesize CMR} & {\footnotesize CRI} &
\multicolumn{1}{l}{\footnotesize NGA} & \multicolumn{1}{l}{} &  &  \\
{\footnotesize 14} & {\footnotesize CPV} & {\footnotesize GNB} &
\multicolumn{1}{l}{\footnotesize RWA} & \multicolumn{1}{l}{} &  &  \\
{\footnotesize 15} & {\footnotesize FIN} & {\footnotesize ISL} &
\multicolumn{1}{l}{\footnotesize TTO} & \multicolumn{1}{l}{} &  &  \\
{\footnotesize 16} & {\footnotesize FJI} & {\footnotesize NAM} &
\multicolumn{1}{l}{\footnotesize PER} & \multicolumn{1}{l}{} &  &  \\
{\footnotesize 17} & {\footnotesize IRN} & {\footnotesize PRT} &
\multicolumn{1}{l}{\footnotesize YUG} & \multicolumn{1}{l}{} &  &  \\
{\footnotesize 18} & {\footnotesize MRT} & {\footnotesize PAK} &
\multicolumn{1}{l}{\footnotesize SOM} & \multicolumn{1}{l}{} &  &  \\
{\footnotesize 19} & {\footnotesize MYS} & {\footnotesize SWZ} &
\multicolumn{1}{l}{\footnotesize TUR} & \multicolumn{1}{l}{} &  &  \\
& \multicolumn{5}{c}{\footnotesize Clusters with two countries} &  \\
{\footnotesize 20} & \multicolumn{1}{r}{\footnotesize ARG} &
\multicolumn{1}{r}{\footnotesize GMB} & {\footnotesize 21} & {\footnotesize %
BEL} & {\footnotesize NOR} &  \\
{\footnotesize 22} & {\footnotesize BRA} & {\footnotesize SUR} &
{\footnotesize 23} & \multicolumn{1}{l}{\footnotesize BWA} &
\multicolumn{1}{l}{\footnotesize MLT} &  \\
{\footnotesize 24} & {\footnotesize CHE} & {\footnotesize USA} &
{\footnotesize 25} & \multicolumn{1}{l}{\footnotesize CHL} &
\multicolumn{1}{l}{\footnotesize GAB} &  \\
{\footnotesize 26} & {\footnotesize COL} & {\footnotesize JAM} &
{\footnotesize 27} & \multicolumn{1}{l}{\footnotesize CYP} &
\multicolumn{1}{l}{\footnotesize SGP} &  \\
{\footnotesize 28} & {\footnotesize DEU} & {\footnotesize FRA} &
{\footnotesize 29} & \multicolumn{1}{l}{\footnotesize DOM} &
\multicolumn{1}{l}{\footnotesize SWE} &  \\
{\footnotesize 30} & {\footnotesize DZA} & {\footnotesize GTM} &
{\footnotesize 31} & \multicolumn{1}{l}{\footnotesize EGY} &
\multicolumn{1}{l}{\footnotesize ZWE} &  \\
{\footnotesize 32} & {\footnotesize GBR} & {\footnotesize NLD} &
{\footnotesize 33} & \multicolumn{1}{l}{\footnotesize HKG} &
\multicolumn{1}{l}{\footnotesize KOR} &  \\
{\footnotesize 34} & {\footnotesize LSO} & {\footnotesize TGO} &
{\footnotesize 35} & \multicolumn{1}{l}{\footnotesize MDG} &
\multicolumn{1}{l}{\footnotesize ZMB} &  \\
{\footnotesize 36} & {\footnotesize MEX} & {\footnotesize URY} &
{\footnotesize 37} & \multicolumn{1}{l}{\footnotesize MOZ} &
\multicolumn{1}{l}{\footnotesize SEN} &  \\
{\footnotesize 38} & {\footnotesize PAN} & {\footnotesize SYR} &
{\footnotesize 39} & \multicolumn{1}{l}{\footnotesize PRY} &
\multicolumn{1}{l}{\footnotesize TUN} &  \\
{\footnotesize 40} & {\footnotesize TCD} & {\footnotesize ZAR} &  &  &  &
\\
& \multicolumn{5}{c}{\footnotesize Two separate countries} &  \\
{\footnotesize 41} & {\footnotesize JPN} & \multicolumn{1}{r}{} &  &  &  &
\\
{\footnotesize 42} & {\footnotesize PHL} & \multicolumn{1}{r}{} &  &  &  &
\\ \hline\hline
\multicolumn{7}{l}{{\footnotesize Listed according to size. }$p_{\min }=0.01$%
{\footnotesize \ and }$l=2.$}%
\end{tabular}%
\end{table}

\begin{table}[tbp]
\caption{Bootstrap: Relative Convergence (PWT)}\centering
\par
\begin{tabular}{rllrrrrr}
\hline\hline
{\footnotesize No} & \multicolumn{2}{l}{\footnotesize Countries} &  &  &  &
&  \\ \hline
{\footnotesize 1} & {\footnotesize COL} & {\footnotesize JAM} &
\multicolumn{1}{l}{\footnotesize MYS} & \multicolumn{1}{l}{\footnotesize NAM}
& \multicolumn{1}{l}{\footnotesize PAN} & \multicolumn{1}{l}{\footnotesize %
SYR} & \multicolumn{1}{l}{\footnotesize TUR} \\
{\footnotesize 2} & {\footnotesize AGO} & {\footnotesize BGD} &
\multicolumn{1}{l}{\footnotesize CMR} & \multicolumn{1}{l}{\footnotesize GHA}
& \multicolumn{1}{l}{\footnotesize MRT} & \multicolumn{1}{l}{\footnotesize %
PAK} & \multicolumn{1}{l}{\footnotesize SOM} \\
{\footnotesize 3} & {\footnotesize CHE} & {\footnotesize CHL} &
\multicolumn{1}{l}{\footnotesize CSK} & \multicolumn{1}{l}{\footnotesize GRC}
& \multicolumn{1}{l}{\footnotesize MEX} & \multicolumn{1}{l}{\footnotesize %
USA} & \multicolumn{1}{l}{} \\
{\footnotesize 4} & {\footnotesize GUY} & {\footnotesize JOR} &
\multicolumn{1}{l}{\footnotesize PNG} & \multicolumn{1}{l}{\footnotesize SLV}
& \multicolumn{1}{l}{\footnotesize SYC} & \multicolumn{1}{l}{\footnotesize %
TUN} & \multicolumn{1}{l}{} \\
{\footnotesize 5} & {\footnotesize CRI} & {\footnotesize GAB} &
\multicolumn{1}{l}{\footnotesize IRN} & \multicolumn{1}{l}{\footnotesize MUS}
& \multicolumn{1}{l}{\footnotesize SUR} & \multicolumn{1}{l}{\footnotesize %
ZAF} & \multicolumn{1}{l}{} \\
{\footnotesize 6} & {\footnotesize CIV} & {\footnotesize COG} &
\multicolumn{1}{l}{\footnotesize LKA} & \multicolumn{1}{l}{\footnotesize MAR}
& \multicolumn{1}{l}{\footnotesize PHL} & \multicolumn{1}{l}{\footnotesize %
THA} & \multicolumn{1}{l}{} \\
{\footnotesize 7} & {\footnotesize AUT} & {\footnotesize FIN} &
\multicolumn{1}{l}{\footnotesize ISL} & \multicolumn{1}{l}{\footnotesize ITA}
& \multicolumn{1}{l}{\footnotesize JPN} & \multicolumn{1}{l}{\footnotesize %
TTO} & \multicolumn{1}{l}{} \\
{\footnotesize 8} & {\footnotesize BDI} & {\footnotesize BUR} &
\multicolumn{1}{l}{\footnotesize GIN} & \multicolumn{1}{l}{\footnotesize HVO}
& \multicolumn{1}{l}{\footnotesize MLI} & \multicolumn{1}{l}{\footnotesize %
MWI} & \multicolumn{1}{l}{} \\
{\footnotesize 9} & {\footnotesize LSO} & {\footnotesize RWA} &
\multicolumn{1}{l}{\footnotesize TCD} & \multicolumn{1}{l}{\footnotesize TGO}
& \multicolumn{1}{l}{\footnotesize ZAR} & \multicolumn{1}{l}{} &
\multicolumn{1}{l}{} \\
{\footnotesize 10} & {\footnotesize CPV} & {\footnotesize GMB} &
\multicolumn{1}{l}{\footnotesize GNB} & \multicolumn{1}{l}{\footnotesize NER}
& \multicolumn{1}{l}{\footnotesize UGA} & \multicolumn{1}{l}{} &
\multicolumn{1}{l}{} \\
{\footnotesize 11} & {\footnotesize DEU} & {\footnotesize DNK} &
\multicolumn{1}{l}{\footnotesize FRA} & \multicolumn{1}{l}{\footnotesize LUX}
& \multicolumn{1}{l}{\footnotesize NZL} & \multicolumn{1}{l}{} &
\multicolumn{1}{l}{} \\
{\footnotesize 12} & {\footnotesize CYP} & {\footnotesize PRT} &
\multicolumn{1}{l}{\footnotesize SGP} & \multicolumn{1}{l}{\footnotesize YUG}
&  &  &  \\
{\footnotesize 13} & {\footnotesize DZA} & {\footnotesize ECU} &
\multicolumn{1}{l}{\footnotesize GTM} & \multicolumn{1}{l}{\footnotesize SWZ}
&  &  &  \\
{\footnotesize 14} & {\footnotesize BOL} & {\footnotesize DOM} &
\multicolumn{1}{l}{\footnotesize KOR} & \multicolumn{1}{l}{\footnotesize PRY}
&  &  &  \\
{\footnotesize 15} & {\footnotesize BRA} & {\footnotesize FJI} &
\multicolumn{1}{l}{\footnotesize MLT} & \multicolumn{1}{l}{\footnotesize PER}
&  &  &  \\
{\footnotesize 16} & {\footnotesize CAF} & {\footnotesize IND} &
\multicolumn{1}{l}{\footnotesize KEN} & \multicolumn{1}{l}{\footnotesize NGA}
&  &  &  \\
{\footnotesize 17} & {\footnotesize BEL} & {\footnotesize GBR} &
\multicolumn{1}{l}{\footnotesize NLD} & \multicolumn{1}{l}{\footnotesize NOR}
&  &  &  \\
{\footnotesize 18} & {\footnotesize BEN} & {\footnotesize MDG} &
\multicolumn{1}{l}{\footnotesize ZMB} & \multicolumn{1}{l}{\footnotesize ZWE}
&  &  &  \\
{\footnotesize 19} & {\footnotesize AUS} & {\footnotesize CAN} &
\multicolumn{1}{l}{\footnotesize SWE} & \multicolumn{1}{l}{} &  &  &  \\
{\footnotesize 20} & {\footnotesize BRB} & {\footnotesize HKG} &
\multicolumn{1}{l}{\footnotesize IRL} & \multicolumn{1}{l}{} &  &  &  \\
{\footnotesize 21} & {\footnotesize BWA} & {\footnotesize MOZ} &
\multicolumn{1}{l}{\footnotesize SEN} & \multicolumn{1}{l}{} &  &  &  \\
{\footnotesize 22} & {\footnotesize ESP} & {\footnotesize ISR} &
\multicolumn{1}{l}{\footnotesize PRI} & \multicolumn{1}{l}{} &  &  &  \\
& \multicolumn{6}{c}{\footnotesize Clusters with two countries} &  \\
{\footnotesize 23} & {\footnotesize HTI} & {\footnotesize IDN} &  &  &  &  &
\\
{\footnotesize 24} & {\footnotesize EGY} & {\footnotesize HND} &  &  &  &  &
\\
{\footnotesize 25} & {\footnotesize ARG} & {\footnotesize URY} &  &  &  &  &
\\
& \multicolumn{6}{c}{\footnotesize One separate country} &  \\
{\footnotesize 26} & {\footnotesize VEN} & \multicolumn{1}{r}{} &  &  &  &
&  \\ \hline\hline
\multicolumn{8}{l}{{\footnotesize Listed according to size. }$p_{\min }=0.01$%
{\footnotesize \ and }$l=2.$} \\
\multicolumn{8}{l}{\footnotesize The number of bootstrap samples is set at
200.}%
\end{tabular}%
\end{table}
\end{center}

\pagebreak
\begin{table}[h]
\caption{NUTS1 code }{\footnotesize
\begin{tabular}{ccclll}
\hline\hline
Code & Country &  & Code & Country &  \\
\multicolumn{1}{l}{} & \multicolumn{1}{l}{} & \multicolumn{1}{l}{} &  &  &
\\
\multicolumn{1}{l}{AT} & \multicolumn{1}{l}{\textit{Austria}} &
\multicolumn{1}{l}{} & IE & \textit{Ireland} &  \\
\multicolumn{1}{l}{AT1} & \multicolumn{1}{l}{} & \multicolumn{1}{l}{
Ostosterreich} &  &  &  \\
\multicolumn{1}{l}{AT2} & \multicolumn{1}{l}{} & \multicolumn{1}{l}{
Sudosterreich} & IT & \textit{Italy} &  \\
\multicolumn{1}{l}{AT3} & \multicolumn{1}{l}{} & \multicolumn{1}{l}{
Westosterreich} & IT1 &  & Nord Ovest \\
\multicolumn{1}{l}{BE} & \multicolumn{1}{l}{\textit{Belgium}} &
\multicolumn{1}{l}{} & IT2 &  & Lombardia \\
\multicolumn{1}{l}{BE1} & \multicolumn{1}{l}{} & \multicolumn{1}{l}{Region
Bruxelles-Capital-Brussels} & IT3 &  & Nord Est \\
\multicolumn{1}{l}{} & \multicolumn{1}{l}{} & \multicolumn{1}{l}{
Hoofdstedelijke Gewest} & IT4 &  & Emilia-Romagna \\
\multicolumn{1}{l}{BE2} & \multicolumn{1}{l}{} & \multicolumn{1}{l}{Vlaams
Gewest} & IT5 &  & Centro \\
\multicolumn{1}{l}{BE3} & \multicolumn{1}{l}{} & \multicolumn{1}{l}{Region
Wallonne} & IT6 &  & Lazio \\
\multicolumn{1}{l}{DE} & \multicolumn{1}{l}{\textit{Germany}} &
\multicolumn{1}{l}{} & IT7 &  & Abruzzo-Molise \\
\multicolumn{1}{l}{DE1} & \multicolumn{1}{l}{} & \multicolumn{1}{l}{
Baden-Wurttemberg} & IT8 &  & Campania \\
\multicolumn{1}{l}{DE2} & \multicolumn{1}{l}{} & \multicolumn{1}{l}{Bayern}
& IT9 &  & Sud \\
\multicolumn{1}{l}{DE3} & \multicolumn{1}{l}{} & \multicolumn{1}{l}{Berlin}
& ITA &  & Sicilia \\
\multicolumn{1}{l}{DE5} & \multicolumn{1}{l}{} & \multicolumn{1}{l}{Bremen}
& ITB &  & Sardegna \\
\multicolumn{1}{l}{DE6} & \multicolumn{1}{l}{} & \multicolumn{1}{l}{Hamburg}
& LU & \textit{Luxembourg} &  \\
\multicolumn{1}{l}{DE7} & \multicolumn{1}{l}{} & \multicolumn{1}{l}{Hessen}
&  &  &  \\
\multicolumn{1}{l}{DE9} & \multicolumn{1}{l}{} & \multicolumn{1}{l}{
Niedersachsen} & NL & \textit{Netherlands} &  \\
\multicolumn{1}{l}{DEA} & \multicolumn{1}{l}{} & \multicolumn{1}{l}{
Nordrhein-Westfalen} & NL1 &  & Noord-Nederland \\
\multicolumn{1}{l}{DEB} & \multicolumn{1}{l}{} & \multicolumn{1}{l}{
Rheinland-Pfalz} & NL2 &  & Oost-Nederland \\
\multicolumn{1}{l}{DEC} & \multicolumn{1}{l}{} & \multicolumn{1}{l}{Saarland}
& NL3 &  & West-Nederland \\
\multicolumn{1}{l}{DEG} & \multicolumn{1}{l}{} & \multicolumn{1}{l}{Thuringen
} & NL4 &  & Zuid-Nederland \\
\multicolumn{1}{l}{DK} & \multicolumn{1}{l}{\textit{Denmark}} &
\multicolumn{1}{l}{} & PT & \textit{Portugal} &  \\
&  &  & PT1 &  & Continente \\
\multicolumn{1}{l}{ES} & \multicolumn{1}{l}{\textit{Spain}} &
\multicolumn{1}{l}{} & SE & \textit{Sweden} &  \\
\multicolumn{1}{l}{ES3} & \multicolumn{1}{l}{} & \multicolumn{1}{l}{
Comunidad de Madrid} &  &  &  \\
\multicolumn{1}{l}{ES4} & \multicolumn{1}{l}{} & \multicolumn{1}{l}{Centro}
& UK & \textit{United Kingdom} &  \\
\multicolumn{1}{l}{ES5} & \multicolumn{1}{l}{} & \multicolumn{1}{l}{Este} &
UKC &  & North East \\
\multicolumn{1}{l}{ES6} & \multicolumn{1}{l}{} & \multicolumn{1}{l}{Sur} &
UKD &  & North West \\
\multicolumn{1}{l}{ES7} & \multicolumn{1}{l}{} & \multicolumn{1}{l}{Canarias}
& UKE &  & Yorkshire and \\
\multicolumn{1}{l}{F1} & \multicolumn{1}{l}{\textit{Finland}} &
\multicolumn{1}{l}{} &  &  & Humber \\
&  &  & UKF &  & East Midland \\
\multicolumn{1}{l}{FR} & \multicolumn{1}{l}{\textit{France}} &
\multicolumn{1}{l}{} & UKG &  & West Midlands \\
\multicolumn{1}{l}{FR1} & \multicolumn{1}{l}{} & \multicolumn{1}{l}{Ile de
France} & UKH &  & East of England \\
\multicolumn{1}{l}{FR2} & \multicolumn{1}{l}{} & \multicolumn{1}{l}{
Bassin-Parisien} & UK1 &  & London \\
\multicolumn{1}{l}{FR3} & \multicolumn{1}{l}{} & \multicolumn{1}{l}{Nord Pas
de Calais} & UKJ &  & South East \\
\multicolumn{1}{l}{FR4} & \multicolumn{1}{l}{} & \multicolumn{1}{l}{Est} &
UKK &  & South West \\
\multicolumn{1}{l}{FR5} & \multicolumn{1}{l}{} & \multicolumn{1}{l}{Ouest} &
UKL &  & Wales \\
\multicolumn{1}{l}{FR6} & \multicolumn{1}{l}{} & \multicolumn{1}{l}{Sud-Ouest
} & UKM &  & Scotland \\
\multicolumn{1}{l}{FR7} & \multicolumn{1}{l}{} & \multicolumn{1}{l}{
Centre-Est} & \multicolumn{1}{c}{} & \multicolumn{1}{c}{} &
\multicolumn{1}{c}{} \\
\multicolumn{1}{l}{FR8} & \multicolumn{1}{l}{} & \multicolumn{1}{l}{
Mediterranee} & \multicolumn{1}{c}{} & \multicolumn{1}{c}{} &
\multicolumn{1}{c}{} \\
\multicolumn{1}{l}{GR} & \multicolumn{1}{l}{\textit{Greece}} &
\multicolumn{1}{l}{} & \multicolumn{1}{c}{} & \multicolumn{1}{c}{} &
\multicolumn{1}{c}{} \\
\multicolumn{1}{l}{GR1} & \multicolumn{1}{l}{} & \multicolumn{1}{l}{Voreia
Ellada} & \multicolumn{1}{c}{} & \multicolumn{1}{c}{} & \multicolumn{1}{c}{}
\\
\multicolumn{1}{l}{GR2} & \multicolumn{1}{l}{} & \multicolumn{1}{l}{Kentriki
Ellada} & \multicolumn{1}{c}{} & \multicolumn{1}{c}{} & \multicolumn{1}{c}{}
\\
\multicolumn{1}{l}{GR3} & \multicolumn{1}{l}{} & \multicolumn{1}{l}{Attiki}
& \multicolumn{1}{c}{} & \multicolumn{1}{c}{} & \multicolumn{1}{c}{} \\
\multicolumn{1}{l}{GR4} & \multicolumn{1}{l}{} & \multicolumn{1}{l}{Nisia
Aigaiou, Kriti} & \multicolumn{1}{c}{} & \multicolumn{1}{c}{} &
\multicolumn{1}{c}{} \\ \hline\hline
\end{tabular}
}
\end{table}

\pagebreak

\begin{table}[h]
\caption{Joint Frequency Distribution}
\label{Table2}\centering
\par
{\footnotesize \ }%
\resizebox{17.5cm}{!}{\begin{tabular}{llllllllllllccc}
\multicolumn{12}{l}{\ \ \ \ \ \ \ \ \ \ \ \ \ \ \ \ \ \ \ \ } &  &  &  \\
\hline\hline
&  &  &  &  &  &  &  &  &  &  &  &  &  &  \\
& \multicolumn{1}{r}{} & \multicolumn{1}{r}{} & \multicolumn{1}{r}{} &
\multicolumn{1}{r}{} & \multicolumn{1}{r}{} & \multicolumn{1}{r}{} &
\multicolumn{1}{r}{} & \multicolumn{1}{r}{} &  &  & \multicolumn{1}{c}{} &
&  &  \\
& \multicolumn{8}{c}{I: Asymptotic} &  &  &  & \multicolumn{3}{c}{} \\
& \multicolumn{8}{c}{Number of Clusters} &  &  &  & \multicolumn{3}{c}{
Summary Statistics} \\
& \multicolumn{1}{r}{} & \multicolumn{1}{r}{} & \multicolumn{1}{r}{} &
\multicolumn{1}{r}{} & \multicolumn{1}{r}{} & \multicolumn{1}{r}{} &
\multicolumn{1}{r}{} & \multicolumn{1}{r}{} &  &  & \multicolumn{1}{c}{} &
&  &  \\
Cluster size & \multicolumn{1}{r}{1} & \multicolumn{1}{r}{2} &
\multicolumn{1}{r}{3} & \multicolumn{1}{r}{4} & \multicolumn{1}{r}{5} &
\multicolumn{1}{r}{6} & \multicolumn{1}{r}{7} & \multicolumn{1}{r}{8} & 9 &
10 & \multicolumn{1}{c}{} &  &  &  \\ \hline
&  &  &  &  &  &  &  &  &  &  & Total Clusters &  &  &  \\
Agriculture & \multicolumn{1}{r}{0} & \multicolumn{1}{r}{3} &
\multicolumn{1}{r}{7} & \multicolumn{1}{r}{2} & \multicolumn{1}{r}{4} &
\multicolumn{1}{r}{1} & \multicolumn{1}{r}{1} & \multicolumn{1}{r}{0} & 0 & 0
& \multicolumn{1}{c}{18} &  &  &  \\
Manufacturing & \multicolumn{1}{r}{0} & \multicolumn{1}{r}{7} &
\multicolumn{1}{r}{9} & \multicolumn{1}{r}{4} & \multicolumn{1}{r}{1} &
\multicolumn{1}{r}{1} & \multicolumn{1}{r}{0} & \multicolumn{1}{r}{0} & 0 & 0
& \multicolumn{1}{c}{22} &  &  &  \\
Market Service & \multicolumn{1}{r}{1} & \multicolumn{1}{r}{9} &
\multicolumn{1}{r}{3} & \multicolumn{1}{r}{6} & \multicolumn{1}{r}{0} &
\multicolumn{1}{r}{0} & \multicolumn{1}{r}{1} & \multicolumn{1}{r}{0} & 1 & 0
& \multicolumn{1}{c}{21} &  &  &  \\
Non-market Service & \multicolumn{1}{r}{1} & \multicolumn{1}{r}{6} &
\multicolumn{1}{r}{7} & \multicolumn{1}{r}{2} & \multicolumn{1}{r}{1} &
\multicolumn{1}{r}{1} & \multicolumn{1}{r}{1} & \multicolumn{1}{r}{1} & 0 & 0
& \multicolumn{1}{c}{20} &  &  &  \\
& \multicolumn{1}{r}{} & \multicolumn{1}{r}{} & \multicolumn{1}{r}{} &
\multicolumn{1}{r}{} & \multicolumn{1}{r}{} & \multicolumn{1}{r}{} &
\multicolumn{1}{r}{} & \multicolumn{1}{r}{} &  &  & \multicolumn{1}{c}{} & $%
\sigma _{\bar{y}}$ \ \  & $\bar{y}^{min}$ \  & $\bar{y}^{max}$ \\
Total Clusters & \multicolumn{1}{r}{2} & \multicolumn{1}{r}{25} &
\multicolumn{1}{r}{26} & \multicolumn{1}{r}{14} & \multicolumn{1}{r}{6} &
\multicolumn{1}{r}{3} & \multicolumn{1}{r}{3} & \multicolumn{1}{r}{1} & 1 & 0
& \multicolumn{1}{c}{81} & 15.2 & 9.4 & 103 \\
& \multicolumn{1}{r}{} & \multicolumn{1}{r}{} & \multicolumn{1}{r}{} &
\multicolumn{1}{r}{} & \multicolumn{1}{r}{} & \multicolumn{1}{r}{} &
\multicolumn{1}{r}{} & \multicolumn{1}{r}{} &  &  & \multicolumn{1}{c}{} &
&  &  \\ \hline
& \multicolumn{1}{r}{} & \multicolumn{1}{r}{} & \multicolumn{1}{r}{} &
\multicolumn{1}{r}{} & \multicolumn{1}{r}{} & \multicolumn{1}{r}{} &
\multicolumn{1}{r}{} & \multicolumn{1}{r}{} &  &  & \multicolumn{1}{c}{} &
&  &  \\
& \multicolumn{8}{c}{II: Bootstrap} &  &  &  &  &  &  \\
& \multicolumn{8}{c}{Number of Clusters} &  &  &  & \multicolumn{3}{c}{} \\
& \multicolumn{1}{r}{} & \multicolumn{1}{r}{} & \multicolumn{1}{r}{} &
\multicolumn{1}{r}{} & \multicolumn{1}{r}{} & \multicolumn{1}{r}{} &
\multicolumn{1}{r}{} & \multicolumn{1}{r}{} &  &  & \multicolumn{1}{c}{} &
&  &  \\
Cluster size & \multicolumn{1}{r}{1} & \multicolumn{1}{r}{2} &
\multicolumn{1}{r}{3} & \multicolumn{1}{r}{4} & \multicolumn{1}{r}{5} &
\multicolumn{1}{r}{6} & \multicolumn{1}{r}{7} & \multicolumn{1}{r}{8} & 9 &
10 & \multicolumn{1}{c}{} &  &  &  \\ \hline
& \multicolumn{1}{r}{} & \multicolumn{1}{r}{} & \multicolumn{1}{r}{} &
\multicolumn{1}{r}{} & \multicolumn{1}{r}{} & \multicolumn{1}{r}{} &
\multicolumn{1}{r}{} & \multicolumn{1}{r}{} &  &  & \multicolumn{1}{c}{} &
&  &  \\
Agriculture & \multicolumn{1}{r}{0} & \multicolumn{1}{r}{3} &
\multicolumn{1}{r}{1} & \multicolumn{1}{r}{1} & \multicolumn{1}{r}{1} &
\multicolumn{1}{r}{0} & \multicolumn{1}{r}{1} & \multicolumn{1}{r}{3} & 1 & 1
& \multicolumn{1}{c}{12} &  &  &  \\
Manufacturing & \multicolumn{1}{r}{0} & \multicolumn{1}{r}{2} &
\multicolumn{1}{r}{5} & \multicolumn{1}{r}{1} & \multicolumn{1}{r}{2} &
\multicolumn{1}{r}{3} & \multicolumn{1}{r}{0} & \multicolumn{1}{r}{1} & 1 & 0
& \multicolumn{1}{c}{15} &  &  &  \\
Market Services & \multicolumn{1}{r}{0} & \multicolumn{1}{r}{1} &
\multicolumn{1}{r}{3} & \multicolumn{1}{r}{2} & \multicolumn{1}{r}{2} &
\multicolumn{1}{r}{4} & \multicolumn{1}{r}{1} & \multicolumn{1}{r}{1} & 0 & 0
& \multicolumn{1}{c}{14} &  &  &  \\
Non-market Services & \multicolumn{1}{r}{0} & \multicolumn{1}{r}{1} &
\multicolumn{1}{r}{3} & \multicolumn{1}{r}{2} & \multicolumn{1}{r}{3} &
\multicolumn{1}{r}{3} & \multicolumn{1}{r}{0} & \multicolumn{1}{r}{2} & 0 & 0
& \multicolumn{1}{c}{14} &  &  &  \\
& \multicolumn{1}{r}{} & \multicolumn{1}{r}{} & \multicolumn{1}{r}{} &
\multicolumn{1}{r}{} & \multicolumn{1}{r}{} & \multicolumn{1}{r}{} &
\multicolumn{1}{r}{} & \multicolumn{1}{r}{} &  &  & \multicolumn{1}{c}{} &
&  &  \\
& \multicolumn{1}{r}{} & \multicolumn{1}{r}{} & \multicolumn{1}{r}{} &
\multicolumn{1}{r}{} & \multicolumn{1}{r}{} & \multicolumn{1}{r}{} &
\multicolumn{1}{r}{} & \multicolumn{1}{r}{} &  &  & \multicolumn{1}{c}{} & $%
\sigma _{\bar{y}}$ \ \  & $\bar{y}^{min}$ \  & $\bar{y}^{max}$ \\
Total Clusters & \multicolumn{1}{r}{0} & \multicolumn{1}{r}{7} &
\multicolumn{1}{r}{12} & \multicolumn{1}{r}{6} & \multicolumn{1}{r}{8} &
\multicolumn{1}{r}{10} & \multicolumn{1}{r}{3} & \multicolumn{1}{r}{8} & 2 &
2 & \multicolumn{1}{c}{55} & 5.4 & 11.7 & 62.6 \\
& \multicolumn{1}{r}{} & \multicolumn{1}{r}{} & \multicolumn{1}{r}{} &
\multicolumn{1}{r}{} & \multicolumn{1}{r}{} & \multicolumn{1}{r}{} &
\multicolumn{1}{r}{} & \multicolumn{1}{r}{} &  &  & \multicolumn{1}{c}{} &
&  &  \\ \hline
& \multicolumn{1}{r}{} & \multicolumn{1}{r}{} & \multicolumn{1}{r}{} &
\multicolumn{1}{r}{} & \multicolumn{1}{r}{} & \multicolumn{1}{r}{} &
\multicolumn{1}{r}{} & \multicolumn{1}{r}{} &  &  & \multicolumn{1}{c}{} &
&  &  \\
& \multicolumn{1}{r}{} & \multicolumn{1}{r}{} & \multicolumn{1}{r}{} &
\multicolumn{1}{r}{} & \multicolumn{2}{c}{III} & \multicolumn{1}{r}{} &
\multicolumn{1}{r}{} &  &  & \multicolumn{1}{c}{} &  &  &  \\
\multicolumn{15}{l}{\ \ \ \ \ \ \ \ \ \ \ \ \ \ \ \ \ \ \ \ \ \ \ \ \ \ \
Correlation Between Asymptotic and Bootstrap Cluster Outcomes} \\
&  &  &  &  &  &  &  &  &  &  &  &  &  &  \\
Agriculture & \multicolumn{10}{c}{0.672} & \multicolumn{1}{c}{} &  &  &  \\
$\text{Manufacturing}$ & \multicolumn{10}{c}{0.509} & \multicolumn{1}{c}{} &
&  &  \\
$\text{Market Services}$ & \multicolumn{10}{c}{0.557} & \multicolumn{1}{c}{}
&  &  &  \\
$\text{Non-Market Services}$ & \multicolumn{10}{c}{0.591} &
\multicolumn{1}{c}{} &  &  &  \\ \hline
\multicolumn{15}{l}{NB: $\sigma _{\bar{y}}$ denotes the standard deviation
of cluster means. $\bar{y}^{min}$ and $\bar{y}^{max}$ denote the Min and Max}
\\
\multicolumn{15}{l}{of cluster means.} \\ \hline\hline
\end{tabular}%
}
\end{table}

\newpage

\begin{figure}[tbph]
\begin{center}
\mbox{
      \subfigure[Relative Convergence in Agriculture: Asymptotic Results]{\scalebox{0.95}{\includegraphics[width=0.5\textwidth]{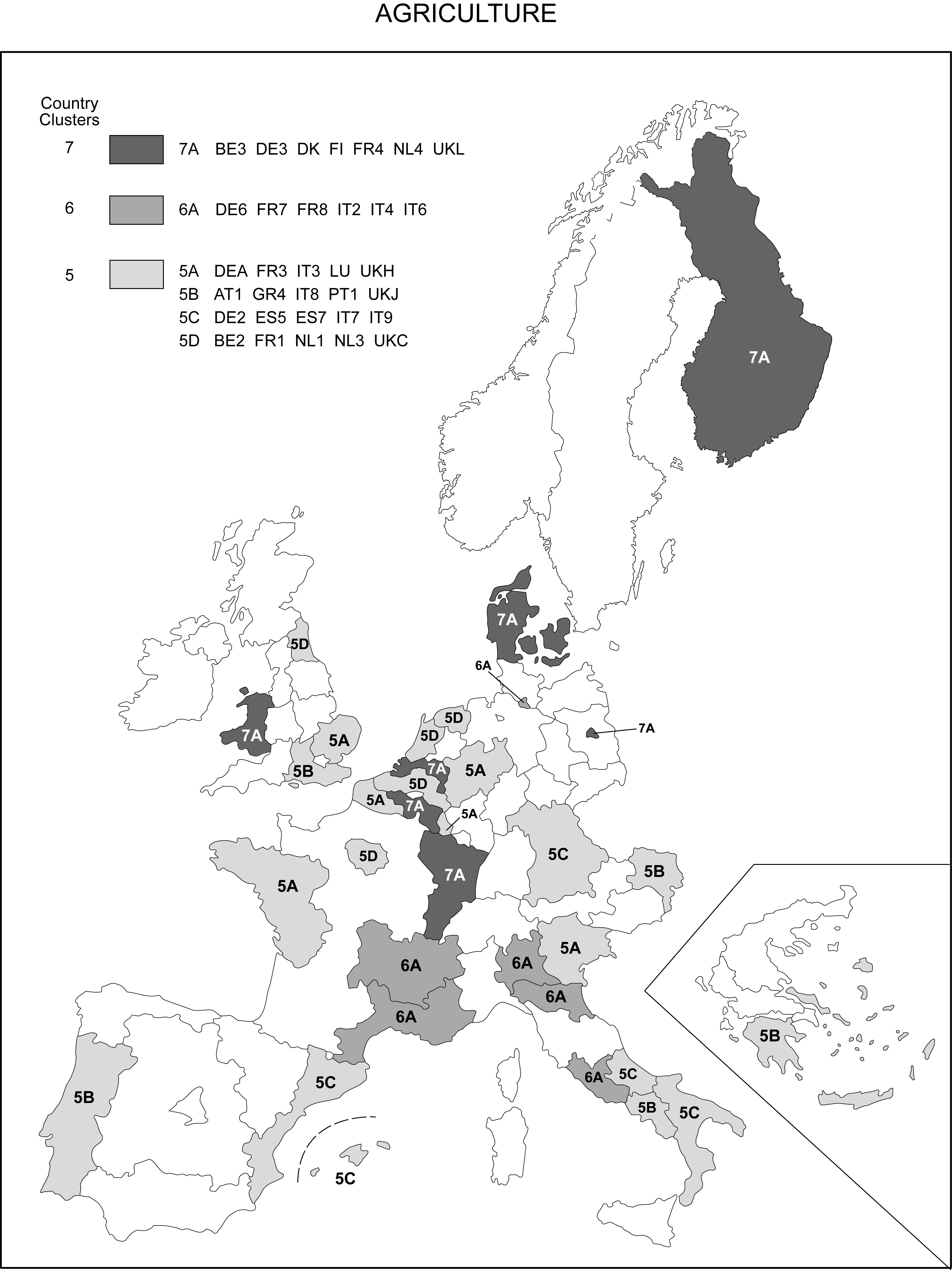}}} \quad
      \subfigure[Relative Convergence in Agriculture: Bootstrap Results]{\scalebox{0.95}{\includegraphics[width=0.5\textwidth]{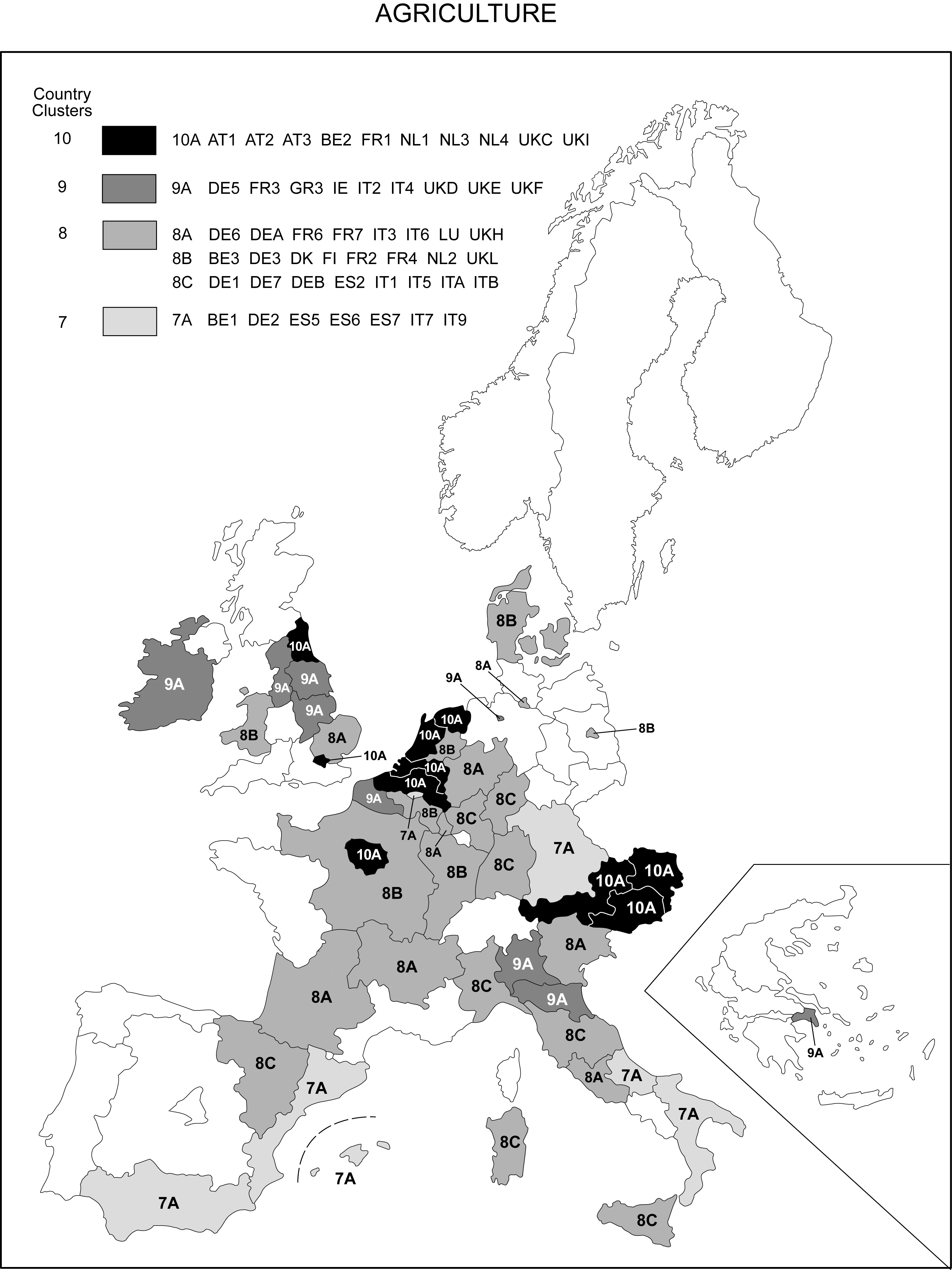}}}
      }
\mbox{
      \subfigure[Relative Convergence in Manufacturing: Asymptotic Results]{\scalebox{0.95}{\includegraphics[width=0.5\textwidth]{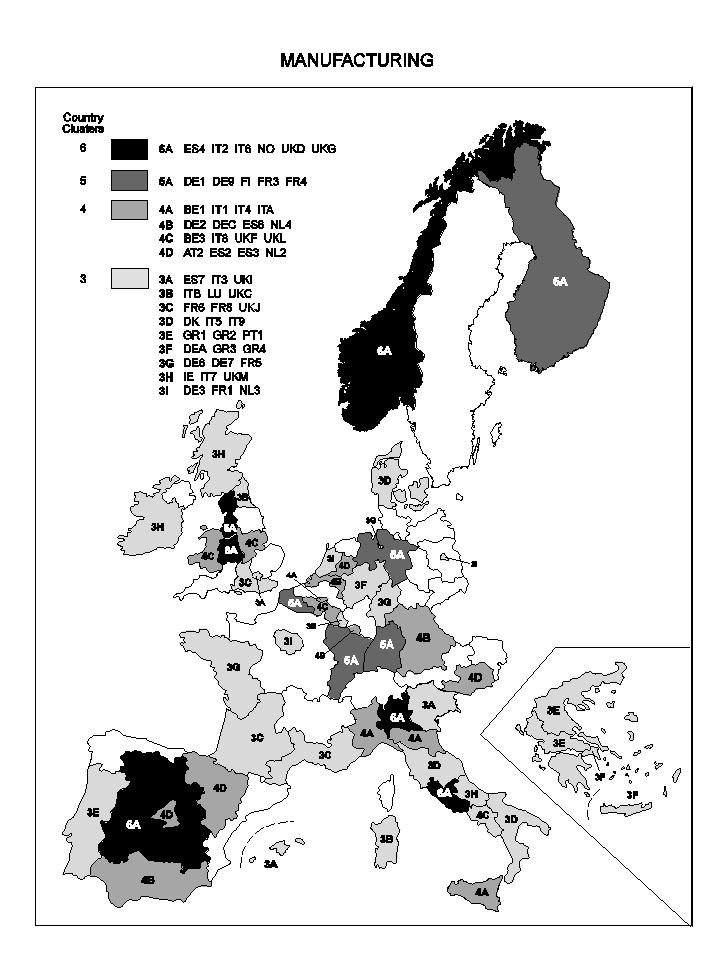}}} \quad
      \subfigure[Relative Convergence in Manufacturing: Bootstrap Results]{\scalebox{0.95}{\includegraphics[width=0.5\textwidth]{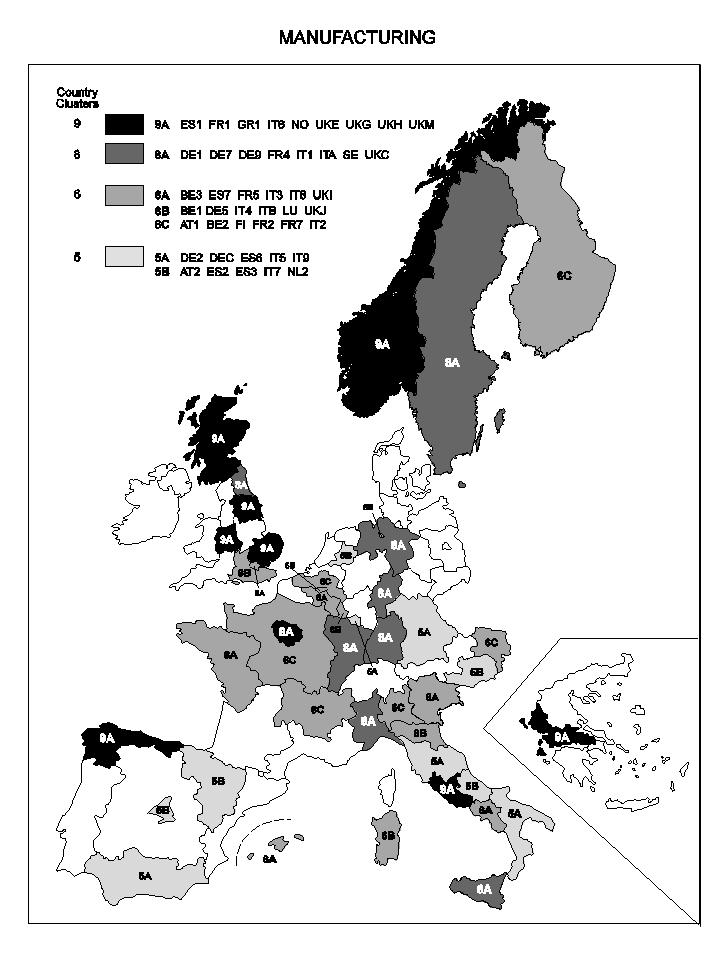}}}
      }
\end{center}
\caption{Asymptotic and Bootstrap Results for Agriculture and Manufacturing}
\label{fig:Results}
\end{figure}

\newpage

\begin{figure}[tbph]
%\begin{figure}[ht]
\par
\begin{center}
\mbox{

      \subfigure[Relative Convergence in Market Services: Asymptotic Results]{\scalebox{0.95}{\includegraphics[width=0.5\textwidth]{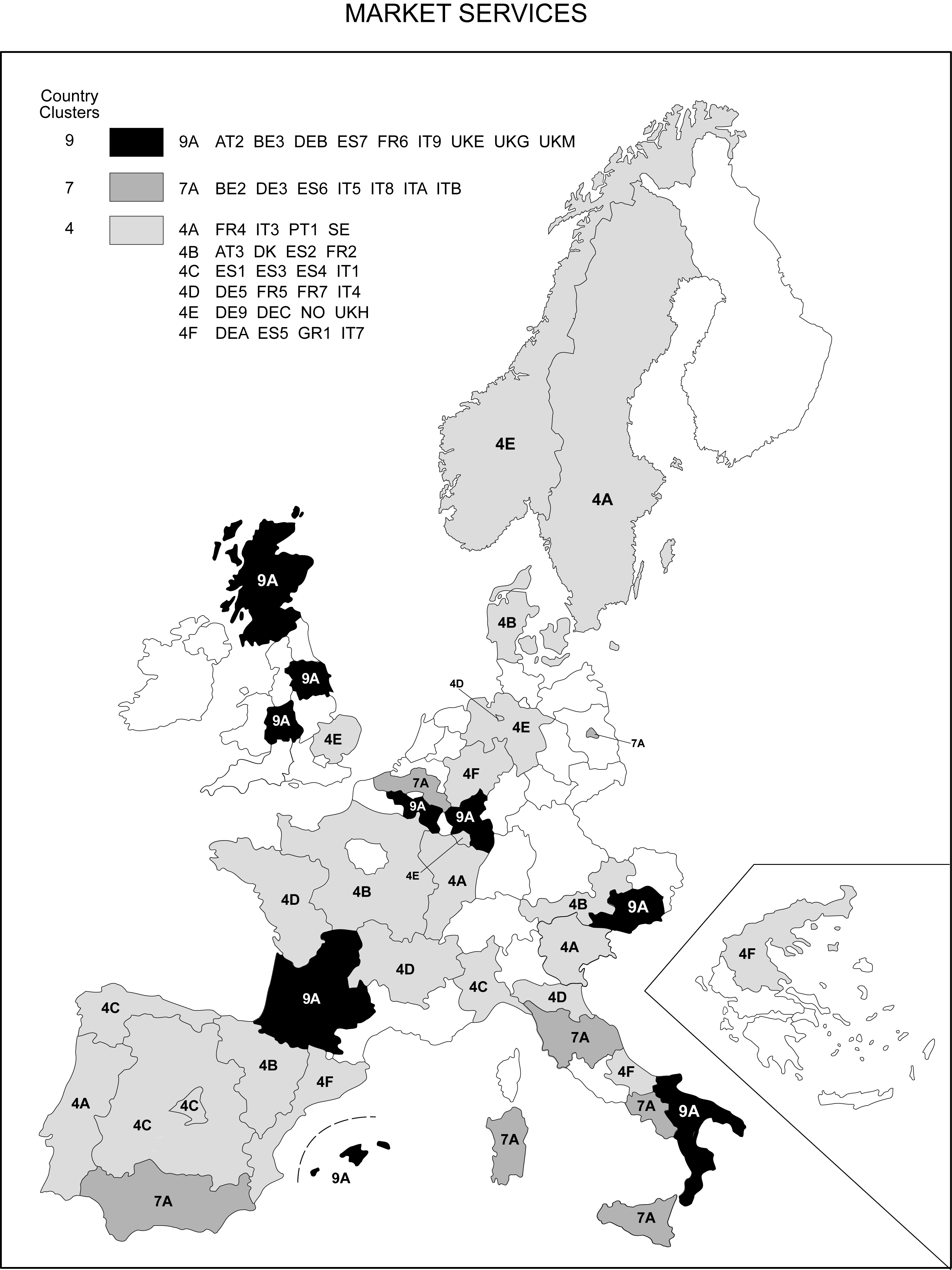}}} \quad
      \subfigure[Relative Convergence in Market Services: Bootstrap Results]{\scalebox{0.95}{\includegraphics[width=0.5\textwidth]{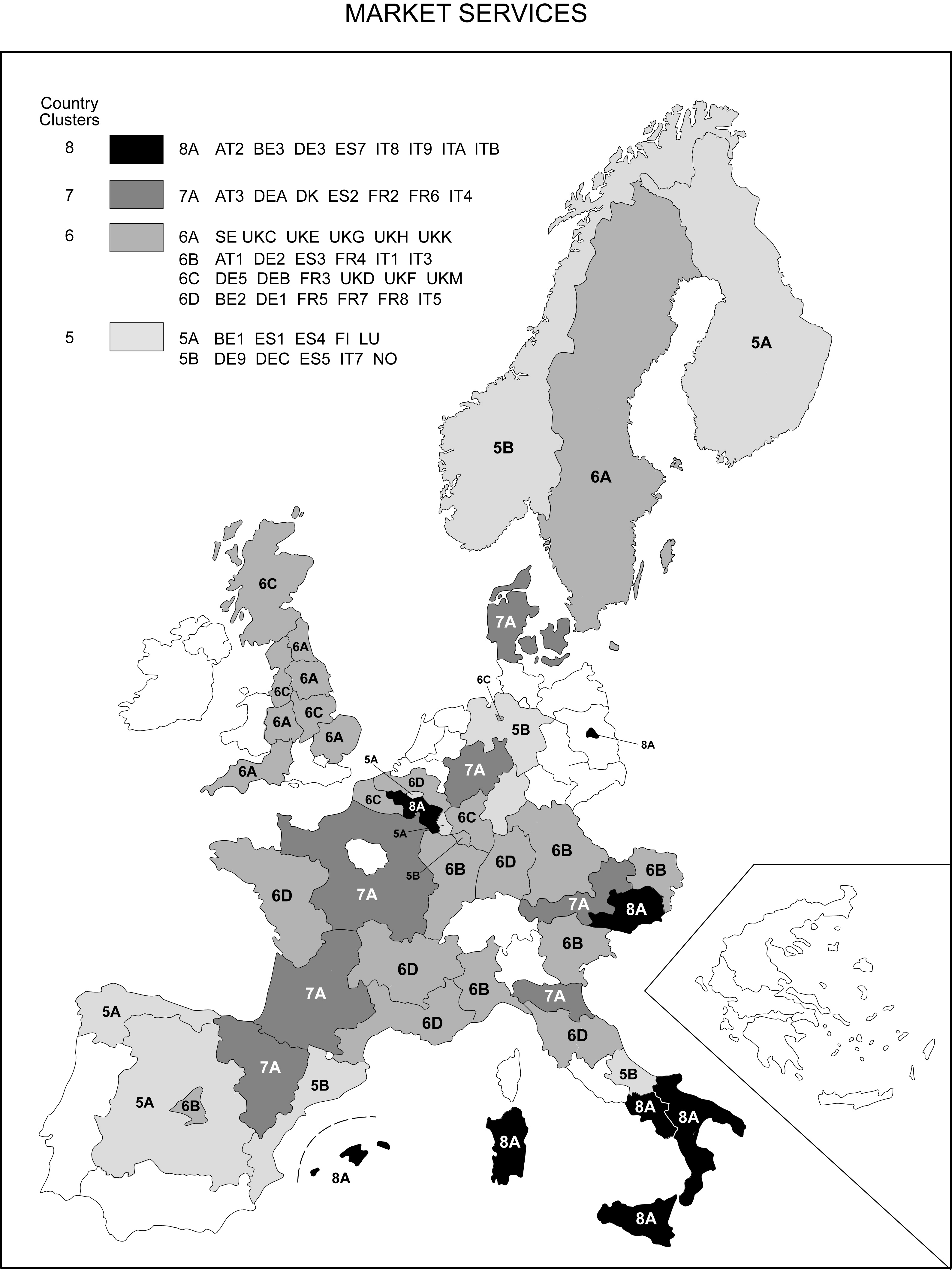}}}
      }
\mbox{
       \subfigure[Relative Convergence in Non-Market Services: Asymptotic Results]{\scalebox{0.95}{\includegraphics[width=0.5\textwidth]{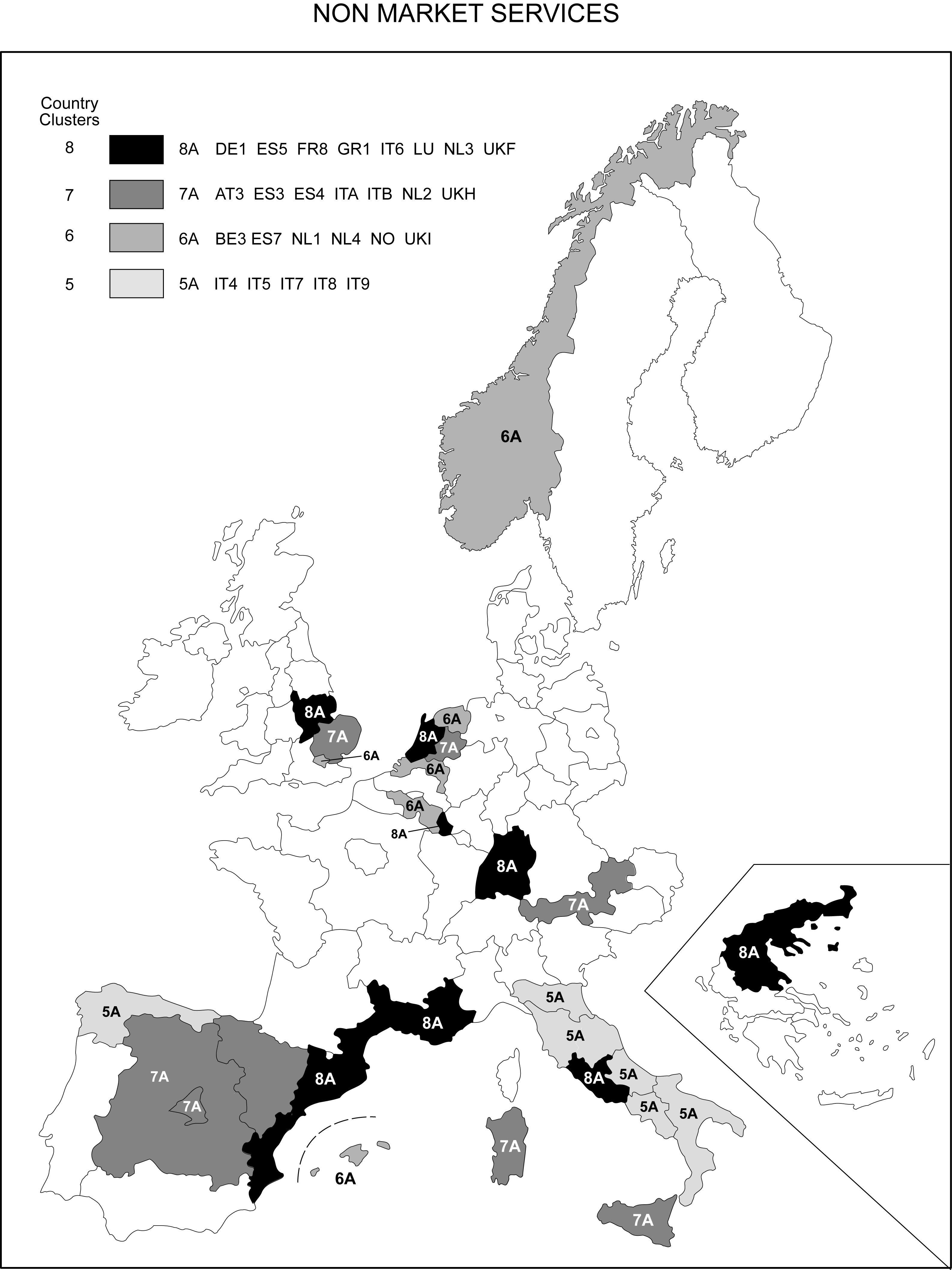}}} \quad
      \subfigure[Relative Convergence in Non-Market Services: Bootstrap Results]{\scalebox{0.95}{\includegraphics[width=0.5\textwidth]{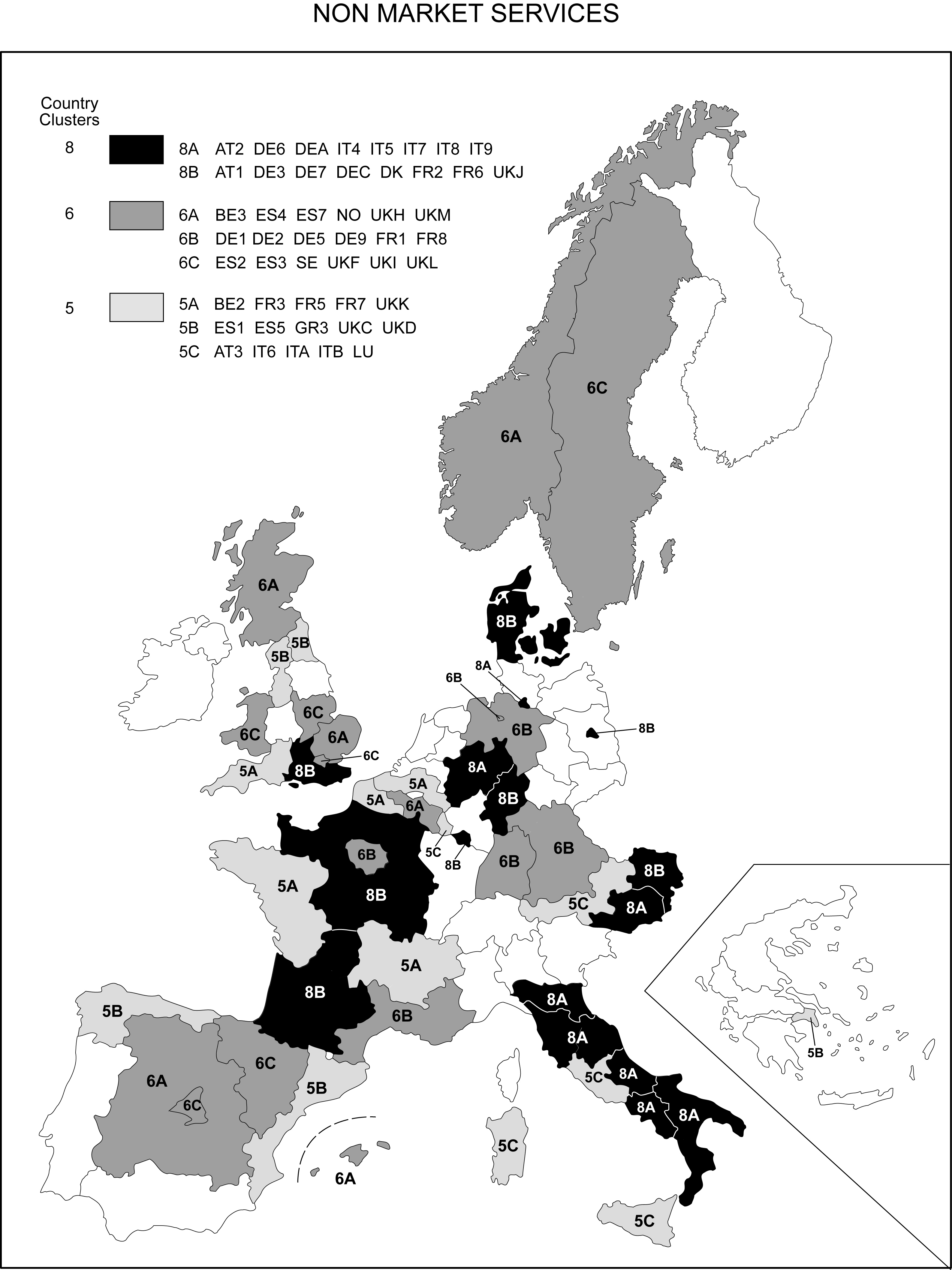}}}
      }
\end{center}
\caption{Asymptotic and Bootstrap Results for Non-Market and Market Services}
\label{fig:Results}
\end{figure}

\begin{figure}[tbph]
\caption{The Distribution of Cluster Size.}
\label{fig:Dist}
\begin{center}
\includegraphics[width=\textwidth]{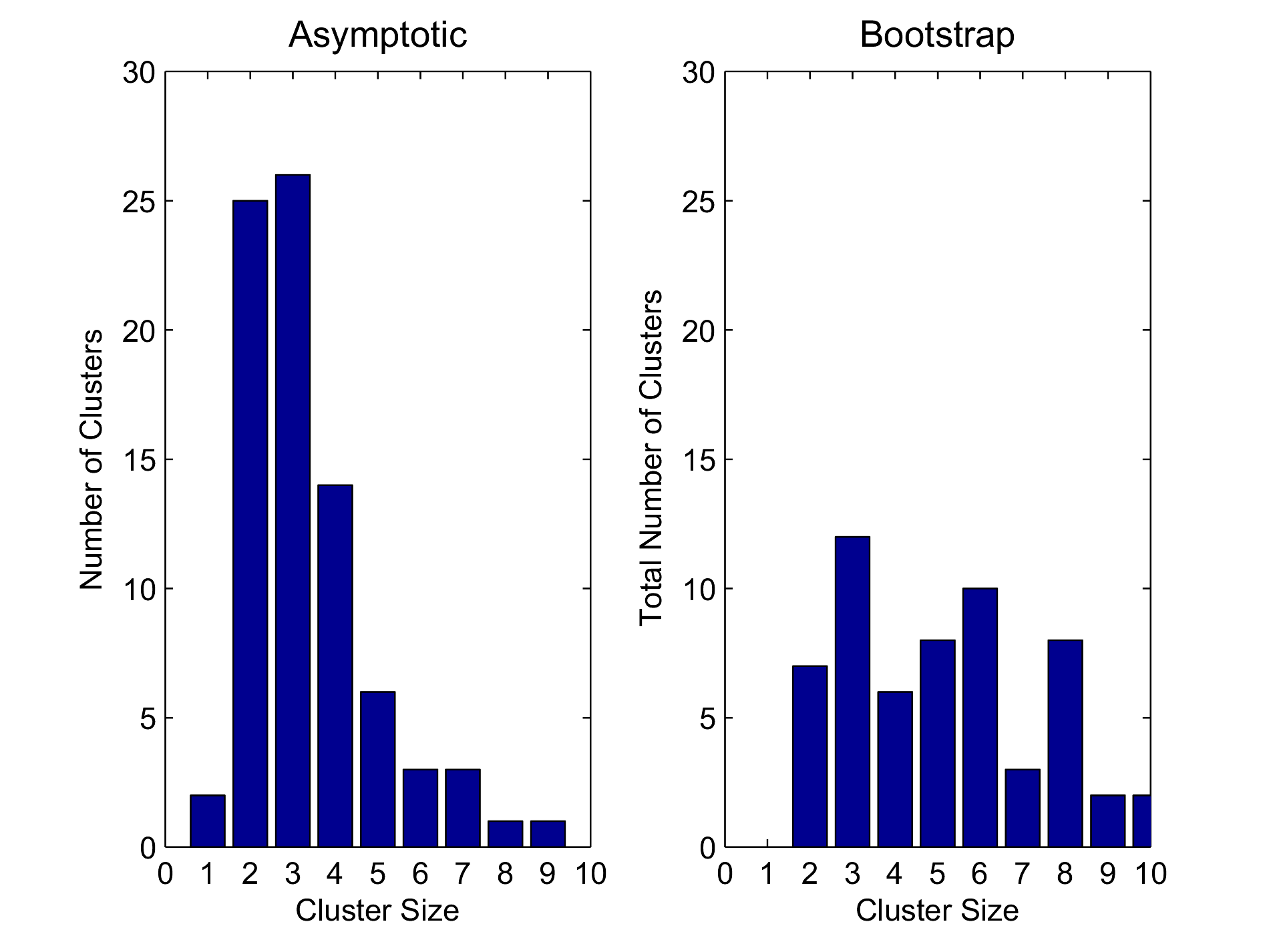}
\par
\includegraphics[width=\textwidth]{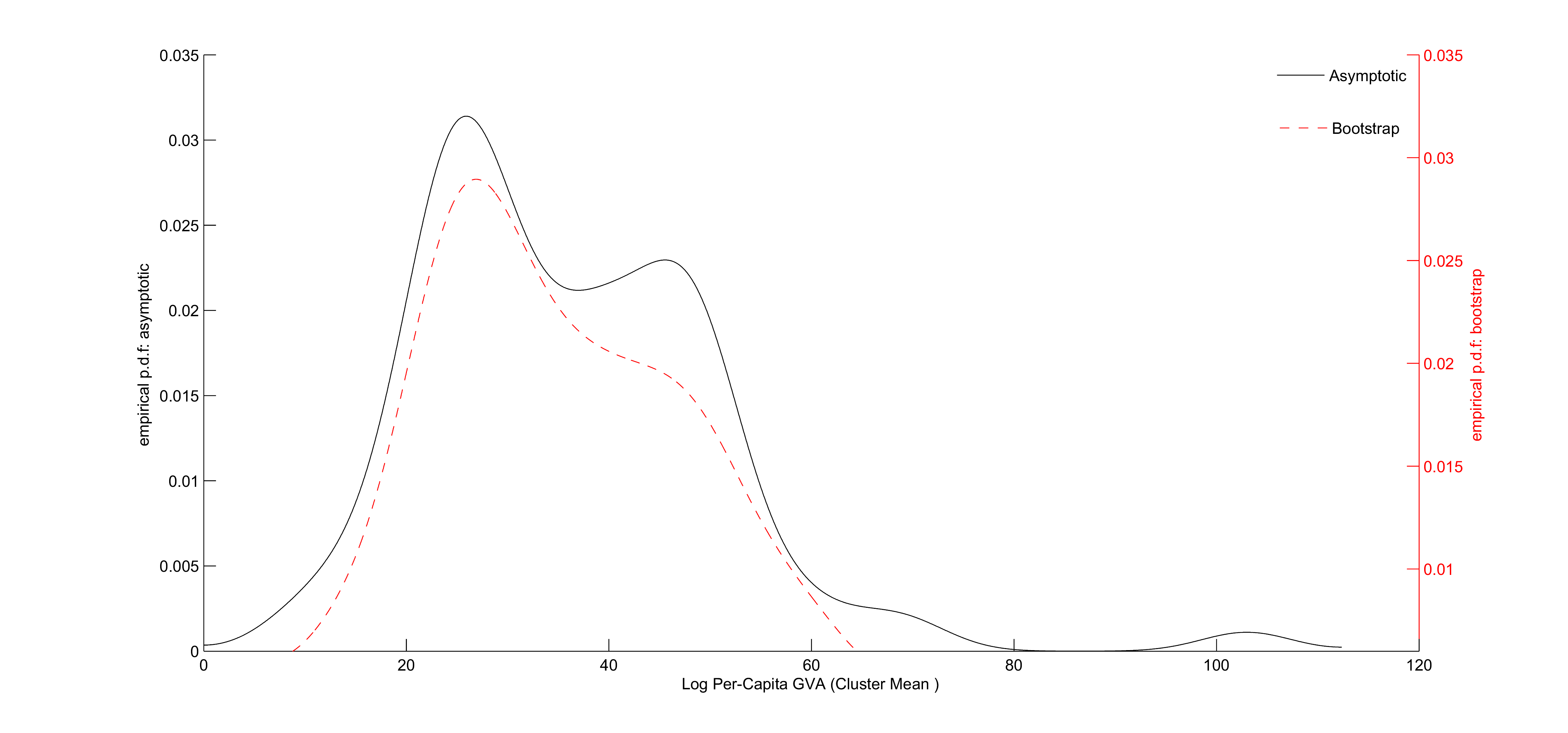}
\end{center}
\caption{The distribution of average log per-capita GVA by cluster: All
sectors. \newline
\newline
Skewness (Asymptotic ) $1.29$ (Bootstrap) $0.27$ \newline
Kurtosis (Asymptotic ) $6.82$ (Bootstrap) $2.20$ }
\end{figure}

\end{document}